\documentclass[preprint,aps]{revtex4-1}
\usepackage{color}
\usepackage{soul}
\usepackage{graphicx, bm, amssymb}
\usepackage[fleqn]{amsmath}
\usepackage{hyperref, cleveref}
\usepackage{setspace, float}
\usepackage{epstopdf}
\usepackage{color}
\usepackage{soul}
\usepackage{array}
\usepackage{multirow}
\usepackage[toc,page]{appendix}

\begin{document}
\def\be{\begin{equation}}
\def\ee{\end{equation}}
\def\ba{\begin{array}}
\def\ea{\end{array}}
\def\bea{\begin{eqnarray}}
\def\eea{\end{eqnarray}}
\def\bc{\begin{center}}
\def\ec{\end{center}}

\title{Twist-2 Pseudoscalar and Vector Meson Distribution Amplitudes in Light-Front Quark Model
with Exponential-type Confining Potential}
\author{Nisha Dhiman$^{a}$, Harleen Dahiya$^{a}$, Chueng-Ryong Ji$^{b}$, and Ho-Meoyng Choi$^{c}$}
\affiliation
{
{$^{a}$Department of Physics, Dr. B. R. Ambedkar National Institute of Technology, Jalandhar-144011, India}\\
{$^{b}$Department of Physics, North Carolina State University, Raleigh, North Carolina 27695-8202, USA}\\
{$^{c}$Department of Physics, Teachers College, Kyungpook National University, Daegu, Korea { 41566}}
}

\begin{abstract}
We study the twist-2 distribution amplitudes (DAs) and the decay constants of pseudoscalar light ($\pi$, $K$) and heavy ($D$, $D_s$, $B$, $B_s$) mesons as well as the  longitudinally and transversely polarized vector light ($\rho$, $K^*$) and heavy ($D^*$, $D_s^*$, $B^*$, $B_s^*$) mesons in the light-front quark model with the Coulomb plus exponential-type confining potential $V_{\rm {exp}} = a + b e^{\alpha r}$ in addition to the hyperfine interaction. We first compute the mass spectra of ground state pseudoscalar and vector light and heavy mesons and fix  the model parameters necessary for the analysis, applying the variational principle with the trial wave function up to the first three lowest order harmonic oscillator (HO) wave functions $\Phi(x, \textbf{k}_\bot) = \sum_{n=1}^{3} c_n \phi_{nS}$. We then obtain the numerical results for the corresponding decay constants of light and heavy mesons. We estimate the DAs, analyze their variation as a function of momentum fraction and compute the first six $\xi$-moments of the $B$ and $D$ mesons as well. We compare our results with the available experimental data as well as with the other theoretical model predictions.
\end{abstract}
\maketitle

\section{Introduction}
\label{sec1}
In the last few decades, the light as well as heavy quark mesonic systems have provided a great deal of important and attractive information on the precise determination of the fundamental parameters of the Standard Model (SM). The nonperturbative structure of the hadron is well described by the hadronic or quark distribution amplitudes (DAs) which not only encode important information on bound states in QCD but also play an essential role in describing the various hard exclusive processes \cite{brodsky1, chernyak1} of QCD via the factorization theorem \cite{collins1} in analogous to parton distributions in inclusive processes. They also help in understanding the distribution of partons in terms of the longitudinal momentum fraction as they are the longitudinal projection of the hadronic wave functions obtained by integrating out the transverse momenta of the fundamental constituents of the hadron \cite{lepage1, efremov1, chernyak2}. Hadronic DAs are defined in terms of vacuum-to-hadron matrix elements of particular non-local quark or quark-gluon operators. The lowest moments of the hadronic DAs for a quark and an antiquark inside a meson provide us the knowledge of decay constants that are considered as direct source of information on the Cabibbo-Kobayashi-Maskawa (CKM) matrix elements, i.e. the fundamental parameters of the SM. The precise determination of decay constants will further allow us to test the unitarity of the quark mixing matrix and CP violation in the SM \cite{mona}. 

\par The $B$-physics phenomenology and the electromagnetic and transition form factors at high $Q^2$ urges the detailed study of hadronic DAs. The predictions of exclusive $B_{u, d, s}$-decays into light pseudoscalar and vector mesons in the context of $CP$ violation and CKM quark mixing matrix require the precise study of $SU (3)$ flavor symmetry breaking effect in the DAs of mesons having strange quark. The hadronic DAs of light mesons were investigated in the pioneering work of Brodsky and Lepage followed by many other studies \cite{lepage1, efremov1, chernyak2, lepage2, efremov2, chernyak3, chernyak4, chernyak5, chernyak6, ali1, braun1, ball1, ball2, ball4, ball5, ball0, ball7, ball8, ball9, ball10}. The hadronic DAs of the heavy $B$-mesons were first investigated by Grozin and Neubert within the heavy quark effective theory (HQET) \cite{grozin1} followed by other studies \cite{kawamura1, lange1, braun2, huang1, ma1, geyer1, khod1, charng1, huang2, lee1, geyer2, yaouanc1, bell1, kawamura2, genon1}. The hadronic DAs of heavy mesons other than $B$-mesons were also discussed in non-HQET framework~\cite{hwang1, choi3, hwang3}. Many theoretical studies using nonperturbative approaches such as the light-front quark model (LFQM) \cite{hwang1, choi3, hwang3, choi4, choi5}, the QCD sum rules (SR) \cite{chernyak2, ball3, ball6, braun3, huang3, huang4, bakulev1, agaev1, mikha1, stefanis1, braguta1, braguta2, braguta3, braguta4}, lattice QCD calculations \cite{khan1, dalley1, braun4, gock1, debbio1}, the chiral-quark model from the instanton vacuum \cite{petrov1, anikin1, nam1, nam2}, the Nambu-Jona-Lasinio (NJL) model \cite{praszal1, arriola1}, the Dyson-Schwinger equation (DSE) approach \cite{chang2, shi1} have motivated the researchers to develop  distinct phenomenological models to estimate the hadronic DAs. Similarly, there have been many theoretical works  using different models such as the LFQM \cite{choi4, choi1, choi2, hwang1, chao1, dhiman1, dhiman2}, the light-front holographic QCD (LFHQCD) \cite{dosch, chang1}, SR \cite{narison, wang1, gel}, the lattice QCD \cite{aoki, bec, na1, na2}, Bethe-Salpeter (BS) model \cite{wang2, cvetic, wang3}, the relativistic quark model (RQM) \cite{hwang2, capstick, ebert}, and the nonrelativistic quark model (NRQM) \cite{yazar1, yazar2, yazar3, shady, hass} that have been devoted to the determination of the decay constants.

\par One of the most successful and efficient nonperturbative approaches is  the LFQM which has been widely used in the phenomenological studies of the hadron physics. It takes advantage of the equal light-front time ($\tau = t + z/c$) quantization and includes the important relativistic effects in the hadronic wave functions \cite{brodsky2, wilson1, dirac} that are neglected in the traditional equal-time Hamiltonian formalism. Apart from having the maximum number (seven) of interaction free (or kinetic) generators, the rational energy-momentum dispersion relation  $p^- = ({\bf{p}}^2_\perp + m^2)/p^+$ yields the sign correlation between the light-front energy $p^- (= p^0 - p^3)$ and the light-front longitudinal momentum $p^+ (= p^0 + p^3)$ leading to the suppression of vacuum fluctuations with the decoupling of complicated non-trivial zero modes. Therefore, a clean Fock state expansion of hadronic wave functions based on a simple vacuum can be built \cite{choi6, choi7, brodsky3, brodsky4}. The light-front wave function (LFWF) can be expressed in terms of hadron momentum-independent internal momentum fraction variables making it explicitly Lorentz invariant \cite{brodsky5}. Based on these properties, the LFQM has been developed and successfully applied to evaluate various meson phenomenologies such as the mass spectra of  heavy and light mesons, their decay constants, DAs, form factors, generalized parton distributions (GPDs). {While the direct connection between the LFQM and QCD has still not been established, recent results on the analyses of  twist-2 and twist-3 quark-antiquark DAs for pseudoscalar and vector mesons in LFQM \cite{choi8} indicated that the constituent quark and antiquark could be considered as  dressed constituents including the zero-mode quantum fluctuations from the QCD vacuum. Also, the light-front holography based on the five-dimensional anti-de Sitter (AdS) space-time and the conformal symmetry has helped in understanding the nature of the effective confinement potential and the resulting light-front wave functions for both light and heavy mesons \cite{brodsky6}. The results on the LFQM analysis of the pion form factor in both space-like and time-like regions \cite{choi9} are found to be compatible with the holographic approach to light-front QCD \cite{brodsky7}. These developments  motivate us to  thoroughly analyse the ground state pseudoscalar and vector mesons mass spectra, decay constants and DAs in the LFQM.

\par The ground state mass spectra and the decay constants of pseudoscalar and vector heavy mesons have already been analysed by fixing the model parameters obtained from the linear and harmonic oscillator (HO) confining potentials using the $1S$ state HO wave function in the light-front approach \cite{choi1}. Further, the mass spectra and decay constants of ground state pseudoscalar and vector light and heavy mesons have been computed in the LFQM by fixing the model parameters obtained from the linear confining potential using the two lowest order HO wave functions \cite{choi2}. It is important to mention here that the fixation of model parameters has not been carried in any other potential beyond the linear and HO confining potentials. In view of this, we attempt to model the confining potential into exponential-type, which has been explored in the non-relativistic formulation~\cite{yazar1,yazar2,yazar3,shady,hass}. The trial wave function $\Phi$ will be used for the variational principle to the QCD-motivated Hamiltonian saturating the Fock-state expansion by the constituent quark and antiquark: $H_{q \bar{q}} = H_0 + V_{q \bar{q}}$, where the effective interaction potential $V_{q \bar{q}}$ is the combination of (1) Coulomb plus exponential-type potential and (2) Hyperfine interaction. For our trial wave function, we find that the larger number of HO basis functions ($1S$, $2S$, and $3S$) is required to achieve the efficacy of the model calculations in contrast to the previous analyses of LFQM with the linear and HO confining potentials~\cite{choi1,choi2}, the efficacy of which was already obtained with up to the two lowest order HO wave functions. It is interesting to note in this work that our LFQM predictions are comparable to each other regardless the type of confining potential as far as the efficacy of model prediction is achieved by allowing sufficient number of HO basis functions for the trial wave function. We compare the present LFQM results for the exponential-type confining potential obtained by the trial wave function composed of the three lowest HO basis functions with the previous LFQM results for the linear and HO confining potentials obtained by the trial wave function composed of up to the two lowest HO basis functions~\cite{choi1,choi2}. 
As in the previous work~\cite{choi2}, the optimal model parameters have been obtained by including the hyperfine interaction term and smearing out the Dirac $\delta$ function from it in order to avoid the negative infinity problem. This study can provide important constraints on the future experiments to describe the role of variational parameters.

The present work is focused on the study of the ground state pseudoscalar and vector light ($\pi$, $\rho$, $K$, and $K^*$) and heavy ($D$, $D^*$, $D_s$, $D_s^*$, $B$, $B^*$, $B_s$, $B_s^*$) mesons mass spectra. The decay constants and the twist-2 DAs of pseudoscalar, longitudinally and transversely polarized vector light and heavy mesons have been studied in detail using the LFQM. A comparison has been made for the central potential $V_0 (r)$ versus $r$ for linear, HO and exponential-type potentials. The variations of decay constants of light and heavy pseudoscalar mesons as well as of the longitudinally and transversely polarized  light and heavy vector mesons have also been shown in terms of  the Gaussian parameter $\beta$. Using our optimized model parameters, we have computed the ground state meson mass spectra for pseudoscalar and vector light and heavy mesons. We compare the ground state mass spectra of mesons in the present work (carried out for the three lowest order HO wave functions) with that of the work in Ref. \cite{choi2} (carried out for the two lowest order HO wave functions). We have computed the numerical values of decay constants as well as the ratios of pseudoscalar and vector mesons decay constants ($f_{V}/f_{P}$, $f_{P'}/f_{P}$, and $f_{V'}/f_{V}$) of light and heavy mesons. The curves of normalized DAs have been plotted as a function of momentum fraction $x$ followed by the computation of first six $\xi$-moments. In a nutshell, the purpose of the present work is to calculate the quark DAs of pseudoscalar, longitudinally and transversely polarized vector light and heavy mesons in the  LFQM based on the idea of modelling the potential. This study will not only  provide essential informations on the understanding of the universal nonperturbative quantities 
but also help further analyses of the hard exclusive processes. 

\par The paper is organized as follows: In Sec. \ref{sec2}, we begin with a brief description of the light-front framework followed by the description of our QCD-motivated Hamiltonian. In Sec. \ref{sec3new}, we discuss the procedure of fixing our model parameters through variational principle in our LFQM and present the numerical results of ground state meson mass spectra obtained from the fixed model parameters in comparison with experimental data. In Sec. \ref{sec3}, we present first in Sec. \ref{subsec3-1} the formulae for the quark DAs and decay constants as well as the  $\xi$-moments of pseudoscalar, longitudinally and transversely polarized vector mesons in the LFQM. Then, in Sec. \ref{subsec3-2}, we present the numerical results of the decay constants of pseudoscalar and vector light and heavy mesons. We also compare them with available experimental data and other theoretical model predictions. In the same subsection, we also present our results of the quark DAs for pseudoscalar, longitudinally and transversely polarized vector mesons followed by the $\xi$-moments. The summary and conclusions are given in Sec. \ref{sec5}. In the Appendix, we present the analytic formula of the mass eigenvalues of the ground state pseudoscalar and vector mesons by fixing the model parameters obtained from the exponential-type potential using the mixture of three lowest order HO states as our trial wave function for the variational principle. 

\section{Light-front quark model}
\label{sec2}
We choose to work in the LFQM in which a meson bound state, consisting of a quark $q$ and an antiquark $\bar{q}$ with total momentum $P$ and spin $S$ is represented as \cite{choi6, choi7}
\begin{eqnarray}
\label{eqn:1}
|M(P,S,S_z)\rangle &=& \int\frac{dp_{q}^+d^2\textbf{p}_{q_{\bot}}}{16\pi^3} \frac{dp_{\bar q}^+d^2\textbf{p}_{\bar q_\bot}}{16\pi^3}16\pi^3 \delta^3(\tilde P-\tilde p_{q}-\tilde p_{\bar q}) \nonumber\\ && \times \sum\limits_{\lambda_{q},\lambda_{\bar q}}\Psi^{SS_z}(\tilde p_{q},\tilde p_{\bar q},\lambda_{q},\lambda_{\bar q}) \ 
 |q(p_{q},\lambda_{q})\bar q(p_{\bar q},\lambda_{\bar q})\rangle,
\end{eqnarray}
where $p_{q (\bar q)}$ and $\lambda_{q (\bar q)}$ are the on-mass shell light-front momentum and the light-front helicity of the constituent quark (antiquark), respectively. 
The momentum $\tilde p$ is defined as
\begin{eqnarray}
\label{eqn:2}
\tilde p=(p^+,~\textbf{p}_\perp), \ \textbf{p}_\perp=(p^1,~p^2), \ p^-=\frac{m^2+\textbf{p}_\perp^2}{p^+}, 
\end{eqnarray}
and
\begin{flalign}
\label{eqn:3}
         |q(p_{q},\lambda_{q})\bar q(p_{\bar q},\lambda_{\bar q})\rangle
        &= b^\dagger(p_{q}, \lambda_{q})d^\dagger(p_{\bar q}, \lambda_{\bar q})|0\rangle,\nonumber \\  
        \{b(p', \lambda'),b^\dagger(p, \lambda)\} &=
        \{d(p', \lambda'),d^\dagger(p, \lambda)\} =
        2(2\pi)^3~\delta^3(\tilde p'-\tilde p)~\delta_{\lambda'\lambda}.
\end{flalign}
The light-front momenta $p_{q}$ and $p_{\bar q}$ in terms of light-front variables are defined as
\begin{eqnarray}
\label{eqn:4}
p_{q}^+&=&x_1 P^+, \ \ p_{\bar q}^+=x_2 P^+,\nonumber \\
\textbf{p}_{q_{\perp}}&=&x_1\textbf{P}_{\perp}+\textbf{k}_{\perp}, \ \ \textbf{p}_{\bar{q}_\perp}=x_2\textbf{P}_{\perp}-\textbf{k}_{\perp},
\end{eqnarray}
where $x_{1 (2)}$ is the longitudinal momentum fraction satisfying the relation $x_1 + x_2 = 1$ and $\textbf{k}_{\perp}$ is the relative transverse momentum of the constituent.\\
The momentum-space light-front wave function $\Psi^{SS_z}$ in Eq. (\ref{eqn:1}) can be expressed as a covariant form
\begin{eqnarray}
\label{eqn:5}
        \Psi^{SS_z}(\tilde p_{q},\tilde p_{\bar q},\lambda_{q},\lambda_{\bar q})
                ={\sqrt{p_{q}^+p_{\bar q}^+}\over \sqrt{2} ~{\sqrt{{M_0^2} - (m_{q} - m_{\bar q})^2}}}
        ~\bar u(p_{q},\lambda_{q})\,\Gamma\, v(p_{\bar q},\lambda_{\bar q})~\sqrt{\partial{k_z}\over \partial{x}}~\Phi(x, \textbf{k}_\bot),\nonumber\\
\end{eqnarray}
where $\Phi(x, \textbf{k}_\bot)$ describes the momentum distribution of the constituents in the bound state with $x \equiv x_1$ and
\begin{eqnarray}
\label{eqn:6}
M^2_0&=& \frac{m^2_{q}+\textbf{k}_\bot^2}{x_1}
         + \frac{m^2_{\bar q}+\textbf{k}_\bot^2}{x_2},
\end{eqnarray}
is the invariant mass squared of the $q \bar{q}$ system. We note that $M_0$ is generally different from the mass $M$ of the meson because the meson, 
quark and antiquark cannot be simultaneously on mass-shell.
Also, the vertex factors $\Gamma$ for pseudoscalar ($\Gamma_P$) and vector ($\Gamma_V$) mesons are given by
\begin{eqnarray}
\label{eqn:7}
 \Gamma_P&=&\gamma_5, \nonumber \\
 \Gamma_V&=&-\not{\! \hat{\varepsilon}}(S_z)+{\hat{\varepsilon}\cdot(p_{q}-p_{\bar q})
                \over M_0+m_{q}+m_{\bar q}},
\end{eqnarray}
with
\begin{eqnarray}
\label{eqn:8}
        &&	\hat{\varepsilon}^\mu(\pm 1) =
                \left[{2\over P^+} \vec \varepsilon_\bot (\pm 1) \cdot
                \vec P_\bot,\,0,\,\vec \varepsilon_\bot (\pm 1)\right], \nonumber\\
            &&   \vec \varepsilon_\bot
                (\pm 1)=\mp(1,\pm i)/\sqrt{2} , \nonumber\\
        &&\hat{\varepsilon}^\mu(0)={1\over M_0}\left({-M_0^2+P_\bot^2\over
                P^+},P^+,P_\bot\right).
\end{eqnarray}
The Dirac spinors satisfy the relations
\begin{eqnarray}
\label{eqn:9}
&&\sum\limits_{\lambda}u(p,\lambda)\bar u(p,\lambda)=\frac{
  \rlap{\hspace{0.03cm}/}{p}+m}{p^+} \,\,\, {\rm for ~ quark} ,\nonumber\\
 &&\sum\limits_{\lambda}v(p,\lambda)\bar v(p,\lambda)=
  \frac{\rlap{\hspace{0.03cm}/}{p}-m}{p^+} \,\,\, {\rm for ~ antiquark} .
\end{eqnarray}
We use the radial wave function $\Phi(x, \textbf{k}_\bot)$ as an expansion of the true wave function in the three lowest order HO 
wave functions $\Phi(x, \textbf{k}_\bot) = \sum_{n=1}^{3} c_n \phi_{nS}$ for both pseudoscalar and vector mesons, respectively. 
The $1S$, $2S$ and $3S$ state HO wave functions are defined as
\begin{eqnarray}
\label{eqn:10}
\phi_{1S}(x,\textbf{k}_\bot)=\frac{1}{(\sqrt{\pi}\beta)^{3/2}}
\exp(-\textbf{k}^{2}/2\beta^{2}),
\end{eqnarray}
\begin{eqnarray}
\label{eqn:11}
\phi_{2S}(x,\textbf{k}_\bot)=\frac{1}{(\sqrt{\pi}\beta)^{3/2}} \bigg(\frac{2 \textbf{k}^{2} - 3 \beta^2}{\sqrt{6} \beta^2}\bigg) \exp(-\textbf{k}^{2}/2\beta^{2}), 
\end{eqnarray}
and
\begin{eqnarray}
\label{eqn:12}
\phi_{3S}(x,\textbf{k}_\bot)=\frac{1}{(\sqrt{\pi}\beta)^{3/2}} \bigg(\frac{15 \beta^4 - 20 \beta^2  \textbf{k}^{2}  + 4 \textbf{k}^{4}}{2 \sqrt{30} \beta^4}\bigg) \exp(-\textbf{k}^{2}/2\beta^{2}),
\end{eqnarray}
where $\beta$ represents the variational parameter and $\textbf{k}^2=\textbf{k}^2_\bot + k^2_z$ is the internal momentum of the meson. The longitudinal component $k_z$ is defined as
\begin{eqnarray}
\label{eqn:13}
k_z &=& \bigg(x-\frac{1}{2}\bigg)M_0 + \frac{m^2_{q}-m^2_{\bar q}}{2M_0}.
\end{eqnarray} 
For the variable transformation $(x, \bf{k}_\bot) \to \bf{k} = (\textbf{k}_\bot, \textit{k}_\textit{z})$, the Jacobian factor $\partial{k_z}/\partial{x}$ is given by
\begin{eqnarray}
\label{eqn:14}
{\partial{k_z}\over \partial{x}} &=& \frac{M_{0}}{4x(1-x)}
\biggl[1-\biggl(\frac{m^2_q-m^2_{\bar{q}}}{M^2_{0}}\biggr)^2 \biggr].
\end{eqnarray} 
The meson wave function can thus be normalized as
\begin{eqnarray}
\label{eqn:15}
        \langle M(P',S',S'_z)|M(P,S,S_z)\rangle = 2(2\pi)^3 P^+
        \delta^3(\tilde P'- \tilde P)\delta_{S'S}\delta_{S'_zS_z}~,
\end{eqnarray}
so that
\begin{eqnarray}
\label{eqn:16}
\int^1_0 dx \int d^2 {\bf k}_\perp~{\partial{k_z}\over \partial{x}}~|\phi_{nS}(x,\textbf{k}_\bot)|^2 = 1. 
\end{eqnarray}
Our LFQM is based on the idea that we consider the radial wave function $\Phi(x, \textbf{k}_\bot)$ as a trial wave function for the variational principle to the QCD-motivated Hamiltonian saturating the Fock-state expansion by the constituent quark and antiquark. In the quark and antiquark center of mass (c.m.) frame, the meson bound system at rest is described by the following QCD-motivated effective Hamiltonian \cite{choi3, choi6, choi7}
\begin{eqnarray}
\label{eqn:17}
H_{\rm c.m.} = \sqrt{\textbf{k}^{2} + m^2_{q}} + \sqrt{\textbf{k}^{2} + m^2_{\bar q}} + V_{q \bar{q}},
\end{eqnarray}
where $V_{q \bar{q}}$ is the effective interaction potential between quark and antiquark in the rest frame of the meson which is given by Coulomb ($V_{\rm Coul}$) plus exponential-type potential ($V_{\rm exp}$) 
in addition to the hyperfine interaction ($V_{\rm hyp}$). That is,
\begin{eqnarray}
\label{eqn:18}
V_{q\bar{q}} &=& V_{0} (r) + V_{\rm hyp} (r)\nonumber\\
&=& V_{\rm exp} + V_{\rm Coul} + V_{\rm hyp}\nonumber\\
&=& a + be^{\alpha r} - \frac{4\kappa}{3r}
+ \frac{2} {3} \frac{\bf{{S}_{q}}\cdot\bf{{S}_{\bar{q}}}}
{m_{q}m_{\bar{q}}}\nabla^{2}V_{\rm Coul},
\end{eqnarray}
where $a$, $b$ and $\alpha$ are the parameters of the potential, $\kappa$ is the strong coupling constant which has been taken as one of the variation parameter in this work, $\langle\bf{{S}_{q}}\cdot\bf{{S}_{\bar{q}}}\rangle$ = $-3/4 \,(1/4)$ for the pseudoscalar (vector) meson, respectively. We note that $\nabla^{2}V_{\rm Coul} = (16\pi \kappa/3)\delta^3(\bf r)$ for the contact interaction, however, we shall smear out  $\delta^3 (\bf r)$ to avoid the negative infinity 
problem~\cite{choi2}.

\par In Fig. \ref{fig:1}, we present the variation of the central potential $V_0(r)$ up to $r=2$ fm used in the present work and compare with other central potentials obtained from the linear and HO confining potentials~\cite{choi1,choi2}. 
As one can see from Fig.~\ref{fig:1} that the three different types of confining potentials are not much different from each other in the relevant range of potential ($r \leq 2$ fm). Nevertheless, these little differences of central potentials may affect the predictions of the ground state meson mass spectra, decay constants and DAs.
\begin{figure}[h]
  \begin{center}
\includegraphics[width=5.0 in]{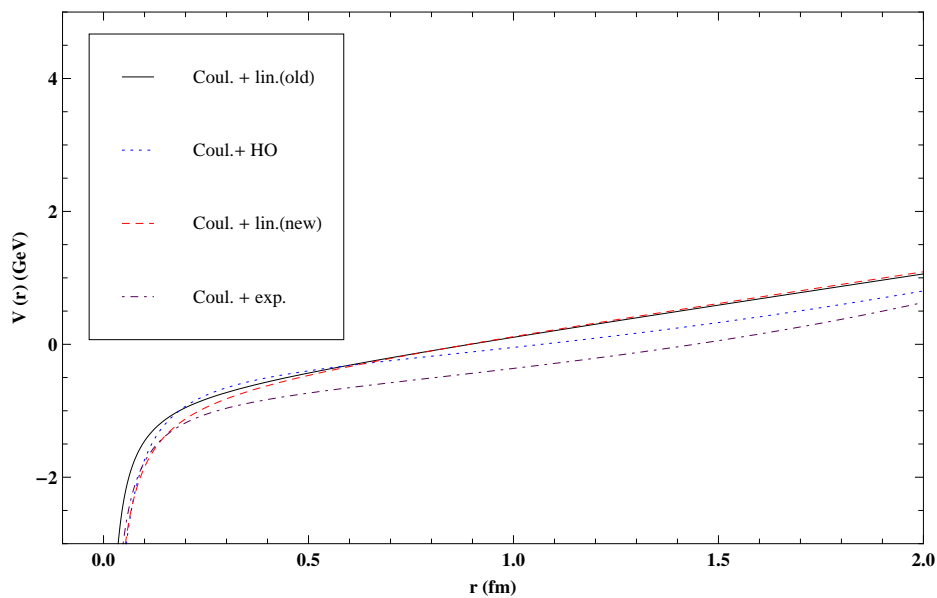}
  \end{center}
  \caption{Variation of central potential $V_0 (r)$ with respect to $r$ in various potential models. Our Coulomb plus exponential-type potential (dotted dashed) is compared with Coulomb plus linear (old CJ Model \cite{choi1}) (solid line), Coulomb plus HO \cite{choi1} (dotted) and Coulomb plus linear (new CJ model \cite{choi2}) (dashed) potentials.} 
  \label{fig:1}
\end{figure}

\section{Fixation of model parameters and Meson Mass Spectra}
\label{sec3new}
\par  We evaluate the expectation value of the system's Hamiltonian $\langle\Phi|H_{\rm c.m.}|\Phi\rangle$ in the variational principle using 
our trial wave function $\Phi(x, \textbf{k}_\bot) = \sum_{n=1}^{3} c_n \phi_{nS}$ consisting of a variational parameter $\beta$. As we discussed in~\cite{choi2}, when we compute $\langle\Phi|H_{\rm c.m.}|\Phi\rangle$ we introduce a Gaussian smearing function which weakens the singularity of $\delta^3 (\bf r)$ in hyperfine interaction, viz., $\delta^3 (\bf r) \to (\sigma^3/\pi^{3/2})\textit{e}^{-\sigma^2 \bf r^2}$, to avoid the negative infinity generated by the $\delta$ function and to find the true minimum value for the mass occuring at a certain value of $\beta$. The analytic formula of mass eigenvalues for our Hamiltonian with the exponential type confining potential, i.e., $M_{q \bar{q}} = \langle\Phi|H_{\rm c.m.}|\Phi\rangle$ is given in the Appendix. The variational principle $M_{q \bar{q}}  / \partial \beta = 0$ gives us a constraint that can be used to rewrite the coupling constant $\kappa$ in terms of other parameters and thus eliminate it from the mass eigenvalues. We then assign a set of values to the externally adjustable variables (through trial and error type of analysis), i.e., ($m_{u(d)}$, $m_s$, $m_c$, $m_b$, $\sigma$, $\alpha$, $c_1$, $c_2$, $c_3$) in order to fix the set of parameters ($a$, $\beta_{q \bar{q}}^P$, $\beta_{q \bar{q}}^V$), where $\beta_{q \bar{q}}^P$ and $\beta_{q \bar{q}}^V$ denote the Gaussian parameters for pseudoscalar and vector mesons, respectively. For the exponential term of confining potential $b e^{{\alpha} r}$, we use a typical value $b=0.385$ GeV reported in Ref.~\cite{yazar1} and find the optimum value of $\alpha$ from the variational principle. Following the same procedure as in Ref.~\cite{choi2}, we use the masses of $\pi$ and $\rho$ as our input values of $M_{q\bar{q}}$. We fix ($a$, $\beta_{q \bar{q}}^P$, $\beta_{q \bar{q}}^V$) for $q=u$, or $d$ from the constraint that the coupling constants $\kappa$ are same for all the ground state pseudoscalar and vector mesons. Effectively, we solve the following three equations:
\begin{eqnarray}
\label{eqn:19}
&&M_{\pi} (\beta_{q \bar{q}}^P, a) = 0.140,\nonumber\\
&&M_{\rho} (\beta_{q \bar{q}}^V, a) = 0.780,\nonumber\\
&&\kappa (\beta_{q \bar{q}}^P, a) = \kappa (\beta_{q \bar{q}}^V, a).
\end{eqnarray}
After solving Eq.~(\ref{eqn:19}), we obtain the values of $a$, $\beta_{q \bar{q}}^P$, $\beta_{q \bar{q}}^V$ as well as the value of $\kappa$. Our obtained value $\kappa=0.4220$ may be in contrast to $\kappa=0.4829$ obtained from the case of linear confining potential~\cite{choi2}. Further, using this common value of $\kappa$, we can then calculate the $\beta$ values for all the other mesons. For the best fit of the meson mass spectra, we obtain $c_1 = \sqrt{0.4}$, $c_2 = \sqrt{0.2}$ and $c_3 = \sqrt{0.4}$. The optimized values of the constituent quark masses and other potential parameters that give the best fit of the ground state mass spectra of mesons are summarized in Table \ref{tab1}.
\begin{table}[h]
\centering
\caption{Constituent quark masses and the potential parameters $\sigma, \alpha$, and $a$ (in units of GeV) obtained by the variational principle for the Hamiltonian with a smeared-out hyperfine interaction. Here $q = u$ and $d$.}
\vspace{0.2cm}
\label{tab1}
\begin{tabular}{m{4.3em}m{4.3em}m{4.3em}m{4.3em}m{4.3em}m{4.3em}m{4.3em}b{4.3em}}
\hline\hline\\[-1.5 ex]
{$m_q$} & {$m_s$} & {$m_c$} & {$m_b$} & {$\sigma$} & {$\alpha$} & {$a$} \\
\hline\\[-1.5 ex]
{0.202} & {0.405} & {1.725} & {5.182} & {0.451} & {0.15} & {$-1.075$} \\[1 ex]
\hline\hline
\end{tabular}
\end{table}
\par Due to the presence of hyperfine interaction in our variational process, we have different sets of $\beta$ values for pseudoscalar and vector mesons, respectively. The optimal Gaussian parameters $\beta_{q \bar{q}}^P$ and $\beta_{q \bar{q}}^V$ for pseudoscalar and vector mesons obtained by the variational principle are listed in Table \ref{tab2}.
\begin{table}[h]
\centering
\caption{The Gaussian parameter $\beta$ (GeV) for ground state pseudoscalar ($\pi$, $K$, $D$, $D_s$, $B$, $B_s$) and vector ($\rho$, $K^*$, $D^*$, $D_s^*$, $B^*$, $B_s^*$) mesons obtained by the variational principle. Here $q = u$ and $d$.}
\vspace{0.2cm}
\label{tab2}
\begin{tabular}{m{5em}m{5em}m{5em}m{5em}m{5em}m{5em}b{5em}}
\hline\hline\\[-1.5 ex]
{$J^{PC}$} & {$\beta_{qq}$} & {$\beta_{qs}$} & {$\beta_{qc}$} & {$\beta_{cs}$} & {$\beta_{qb}$} & {$\beta_{bs}$} \\
\hline\\[-1.5 ex]
{$0^{-+}$} & {0.3387} & {0.2938} & {0.2980} & {0.3010} & {0.3191} & {0.3290} \\[1.5 ex]
{$1^{--}$} & {0.2308} & {0.2437} & {0.2818} & {0.2926} & {0.3115} & {0.3250} \\[1 ex]
\hline\hline
\end{tabular}
\end{table}
 
Using these fixed model parameters, we obtained the ground state pseudoscalar $(K, D_{(s)}, B_{(s)})$ and vector $(K^*, D^*_{(s)}, B^*_{(s)})$ meson mass spectra. Our results are summarized in Table \ref{tab3new}, comparing with the experimental data and the previous results obtained from the linear and HO  potentials~\cite{choi1, choi2}. The predictions for the ground state meson masses in our LFQM obtained from the exponential-type confining potential and smeared hyperfine interaction are in a resonable agreement with the experimental data \cite{tana}. We also note in Table \ref{tab3new} that our predictions are consistent with the ones obtained from the linear \cite{choi1, choi2} and HO \cite{choi1} potential models.

\begin{table}[h]
\centering
\caption{Ground state mass spectra (in units of GeV) of pseudoscalar $(K, D_{(s)}, B_{(s)})$ and vector $(K^*, D^*_{(s)}, B^*_{(s)})$ mesons obtained from the exponential type potential and their comparison with the experimental data~\cite{tana} and the LFQM results obtained 
from the linear and HO potentials~\cite{choi1,choi2}.}
\vspace{0.2cm}
\label{tab3new}
\begin{tabular}{lclclclclcl}
\hline\hline\\[-1.5 ex]
\multicolumn{1}{c}{} &\hspace{0.1in} {$M_{K}$} & \multicolumn{1}{c}{$M_{K^*}$} & {$M_{D}$} & \multicolumn{1}{c}{$M_{D^*}$} & {$M_{D_s}$} & \multicolumn{1}{c}{$M_{D_s^*}$} & {$M_{B}$} & \multicolumn{1}{c}{$M_{B^*}$} & {$M_{B_s}$} & \multicolumn{1}{c}{$M_{B_s^*}$}
\\[1 ex]
\hline\\[-1.5 ex]
Present work                                                                                   &\hspace{0.1in} 0.521 &\hspace{0.05in} 0.826 &\hspace{0.05in} 1.803 &\hspace{0.05in} 1.884 &\hspace{0.05in} 1.929 &\hspace{0.05in} 1.971 &\hspace{0.05in} 5.212 &\hspace{0.05in} 5.242 &\hspace{0.05in} 5.313 &\hspace{0.05in} 5.329
\\[1 ex]
Exp. \cite{tana}                                                                                 &\hspace{0.1in} 0.494 &\hspace{0.05in} 0.892 &\hspace{0.05in} 1.869 &\hspace{0.05in} 2.010 &\hspace{0.05in} 1.968 &\hspace{0.05in} 2.112 &\hspace{0.05in} 5.279 &\hspace{0.05in} 5.325 &\hspace{0.05in} 5.367 &\hspace{0.05in} 5.415
\\[1 ex]
LFQM, Lin \cite{choi1}  
&\hspace{0.1in} 0.478 &\hspace{0.05in} 0.850 &\hspace{0.05in} 1.836 &\hspace{0.05in} 1.998 &\hspace{0.05in} 2.011 &\hspace{0.05in} 2.109 &\hspace{0.05in} 5.235 &\hspace{0.05in} 5.315 &\hspace{0.05in} 5.375 &\hspace{0.05in} 5.424
\\[1 ex]
LFQM, HO \cite{choi1}
&\hspace{0.1in} 0.470 &\hspace{0.05in} 0.875 &\hspace{0.05in} 1.821 &\hspace{0.05in} 2.024 &\hspace{0.05in} 2.005 &\hspace{0.05in} 2.150 &\hspace{0.05in} 5.235 &\hspace{0.05in} 5.349 &\hspace{0.05in} 5.378 &\hspace{0.05in} 5.471
\\[1 ex]
LFQM \cite{choi2} 
&\hspace{0.1in} 0.510 &\hspace{0.05in} 0.835 &\hspace{0.05in} 1.875 &\hspace{0.05in} 1.962 &\hspace{0.05in} 1.981 &\hspace{0.05in} 2.031 &\hspace{0.05in} 5.233 &\hspace{0.05in} 5.268 &\hspace{0.05in} 5.314 &\hspace{0.05in} 5.333
\\[1 ex]
\hline\hline
\end{tabular}
\end{table}
}

\section{Quark Distribution Amplitudes and Decay Constants}
\label{sec3}
Having fixed all the model parameters for the present analysis achieving a reasonable agreement with the data for the meson mass spectra, we now apply the present model calculation to the wave function related observables such as 
the quark DAs and decay constants in this section. We first summarize the relevant formulae in 
Sec.~\ref{subsec3-1} and subsequently present the corresponding numerical results in Sec~\ref{subsec3-2}. 

\subsection{Summary of Formulae}
\label{subsec3-1}

The quark DAs are defined in terms of the matrix elements of non-local operators that are sandwiched between the vacuum and the meson states \cite{hwang1}
\begin{eqnarray}
\label{eqn:20}
\langle 0|\bar{q}(0) \gamma^\mu \gamma_5 q(0)|P (P) \rangle &=& i f_{P}P^{\mu}\int_{0}^{1} \phi_P(x) dx, \\
\label{eqn:21}
\langle 0|\bar{q}(0) \gamma^\mu q(0)|V (P, \, \lambda = 0) \rangle &=& f_{V} M_{V} \epsilon^{\mu}(\lambda)\int_{0}^{1} \phi_{V \parallel}(x) dx, \, \\
\label{eqn:22}
\langle 0|\bar{q}(0) \sigma^{\mu \nu} q(0)|V (P, \, \lambda = \pm 1) \rangle &=& i f_{V}^{\perp}[\epsilon^{\mu}(\lambda) P_{\nu} - \epsilon^{\nu} (\lambda) P_{\mu}]\int_{0}^{1} \phi_{V \perp}(x) dx.
\end{eqnarray}
Here $\phi_P$, $\phi_{V \parallel}$ and $\phi_{V \perp}$ are the twist-2 DAs of pseudoscalar, longitudinally and transversely polarized vector mesons, respectively. 
The explicit forms of quark DAs in our LFQM are given by \cite{choi4}
\begin{eqnarray}
\label{eqn:23}
&&\phi_P(x) = \frac{2\sqrt{6}}{f_P}\int \frac{d^2\textbf{k}_{\perp}}{\sqrt{16 \pi^3}} \sqrt{\partial{k_z}\over \partial{x}}\Phi(x,\textbf{k}_\perp) \frac{\mathcal{A}}{\sqrt{\mathcal{A}^2+\textbf{k}^2_{\perp}}},\\
\label{eqn:24}
&&\phi_{V \parallel}(x) = \frac{2\sqrt{6}}{f_V}\int \frac{d^2\textbf{k}_{\perp}}{\sqrt{16 \pi^3}} \sqrt{\partial{k_z}\over \partial{x}}\frac{\Phi(x,\textbf{k}_\perp)} {{\sqrt{\mathcal{A}^2+\textbf{k}^2_{\perp}}}} \bigg\{\mathcal{A} + \frac{2 \textbf{k}_{\perp}^2}{M_0 + m_q + m_{\bar q}}\bigg\},\\
\label{eqn:25}
&&\phi_{V \perp}(x) = \frac{2\sqrt{6}}{f_V^\perp}\int \frac{d^2\textbf{k}_{\perp}}{\sqrt{16 \pi^3}} \sqrt{\partial{k_z}\over \partial{x}}\frac{\Phi(x,\textbf{k}_\perp)} {{\sqrt{\mathcal{A}^2+\textbf{k}^2_{\perp}}}} \bigg\{\mathcal{A} + \frac{\textbf{k}_{\perp}^2}{M_0 + m_q + m_{\bar q}}\bigg\},
\end{eqnarray}
where $\mathcal{A} = (1 - x) m_q + x m_{\bar q}$.
The quark DAs $\phi=(\phi_P,\phi_{V \parallel},\phi_{V \perp})$ are normalized as 
\begin{eqnarray}
\label{eqn:26}
\int_{0}^{1} \phi(x) dx = 1.
\end{eqnarray}
From Eq.~(\ref{eqn:26}), we can define the decay constants for the pseudoscalar and the vector mesons as
\begin{eqnarray}
\label{eqn:27}
\langle 0|\bar{q} \gamma^\mu \gamma_5 q|P (P) \rangle &=& i f_{P}P^{\mu}, \\
\label{eqn:28}
\langle 0|\bar{q} \gamma^\mu q|V (P, \, \lambda = 0) \rangle &=& f_{V} M_{V} \epsilon^{\mu}(\lambda),\\
\label{eqn:29}
\langle 0|\bar{q} \sigma^{\mu \nu} q|V (P, \, \lambda = \pm 1) \rangle &=& i f_{V}^{\perp}[\epsilon^{\mu}(\lambda) P_{\nu} - \epsilon^{\nu} (\lambda) P_{\mu}].
\end{eqnarray}
We may also define the expectation value of the longitudinal momentum, so-called $\xi$-moments, as follows
\begin{eqnarray}
\label{eqn:30}
\langle\xi^n\rangle = \int_{-1}^{1} d\xi \xi^{n} \phi(\xi) = \int_{0}^{1} dx \xi^{n} \phi(x),
\end{eqnarray}
where $\xi\equiv (1 - x) - x = (1 - 2 x)$.

\subsection{Numerical Results}
\label{subsec3-2}

To perform the numerical calculations of decay constants for pseudoscalar and vector mesons, we use the model parameters given in Table \ref{tab1} and Table \ref{tab2}. However in order to study the sensitivity of our model parameters for the present calculations of the decay constants, we include the systematic errors in our analysis obtained  both from the $\pm 10\%$ variation of $\beta$ values  for the fixed quark masses and  the $\pm 10\%$ variation of quark masses for  the fixed $\beta$ values. As one can see from Tables~\ref{tab4}-\ref{tab9},  the decay constants of our model are more sensitive to the variations of $\beta$ values than those of quark masses.
 
\begin{figure}[h]
\centering
\minipage{0.5\textwidth}
  \includegraphics[width=3 in]{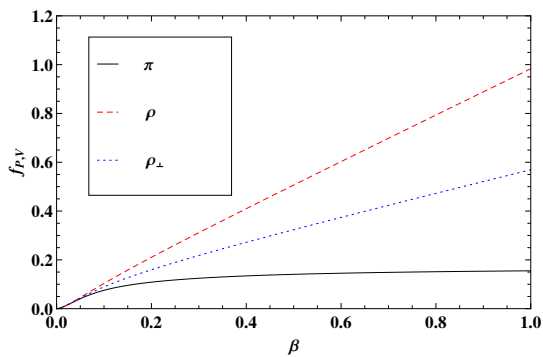}
\endminipage\hfill
\caption{The decay constants of $\pi$, longitudinally and transversely polarized $\rho$ mesons ($f_{\pi}$, $f_{\rho}$ and $f_{\rho}^{\perp}$) 
as functions of the parameter $\beta$ (in GeV).}
\label{fig:2}
\end{figure}

\begin{figure}[h]
\centering
\minipage{0.5\textwidth}
 \includegraphics[width=3 in]{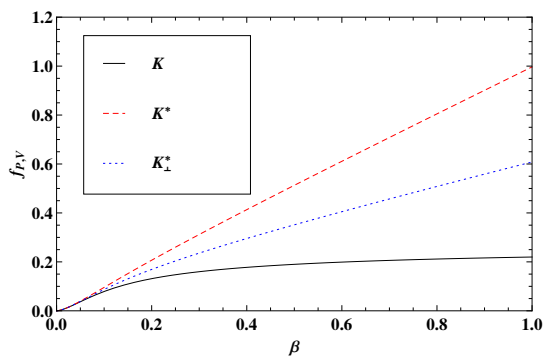}
\endminipage\hfill
\caption{The decay constants $f_{K}$, $f_{K^*}$ and $f_{K^*}^{\perp}$ of pseudoscalar, longitudinally and transversely polarized vector $K$ mesons as functions of the parameter $\beta$ (in GeV).}
\label{fig:3}
\end{figure}

\begin{figure}[h]
\centering
\minipage{0.5\textwidth}
  \includegraphics[width=3 in]{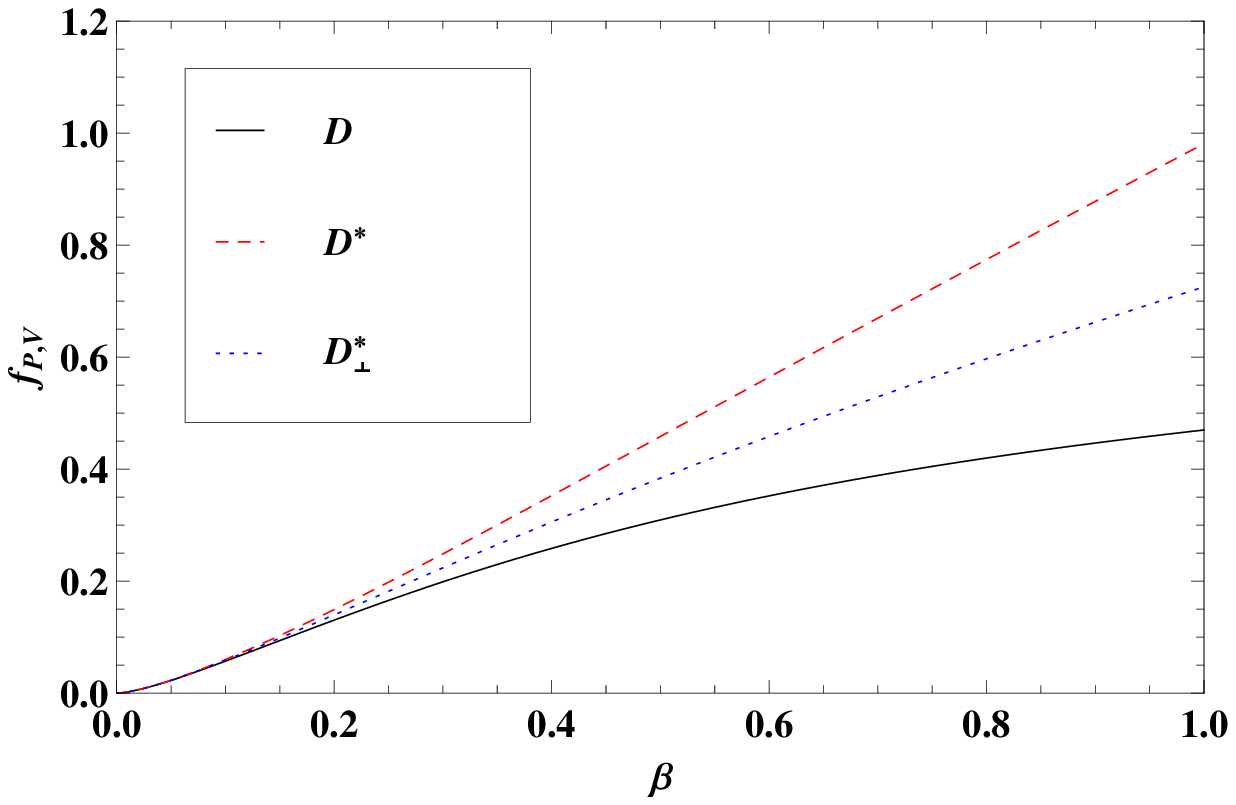}
\endminipage\hfill
\minipage{0.5\textwidth}
 \includegraphics[width=3 in]{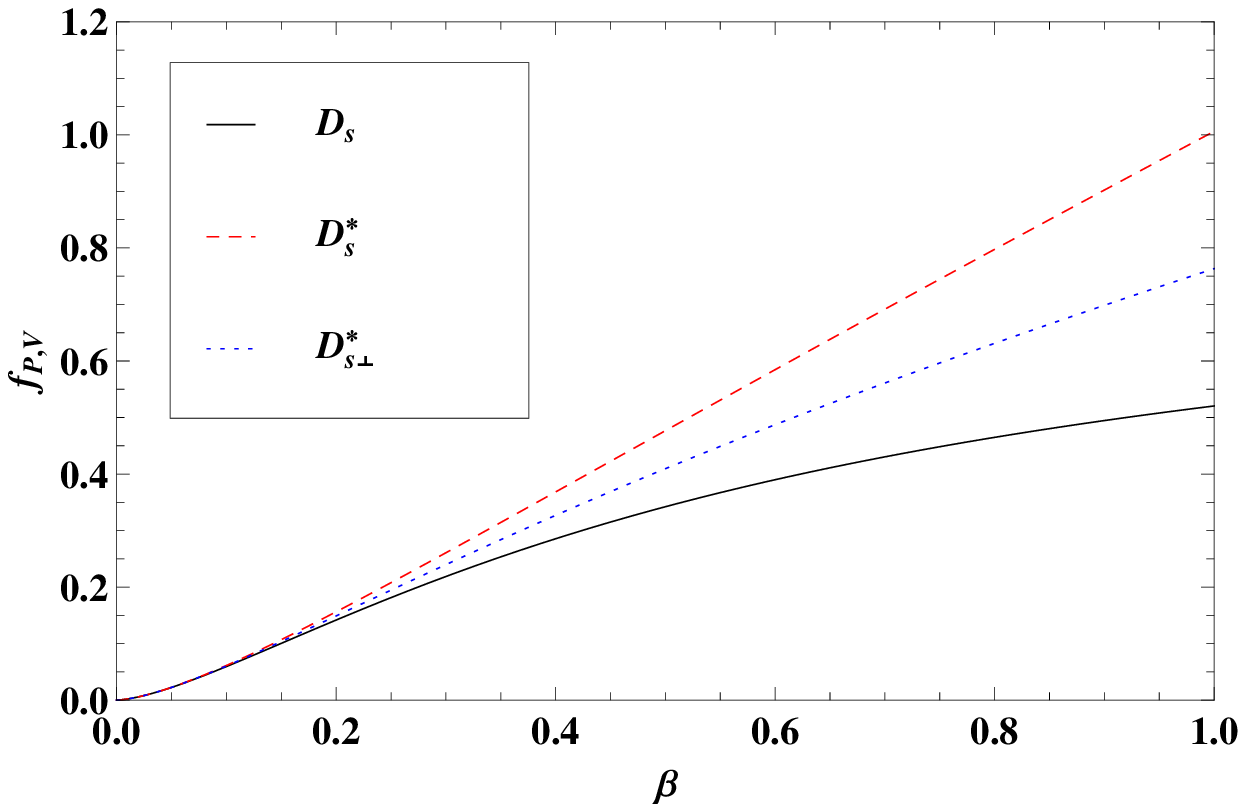}
\endminipage\hfill
\caption{The decay constants $f_{D_{(s)}}$, $f_{D^*_{(s)}}$ and $f_{D^*_{(s)}}^{\perp}$ of pseudoscalar, longitudinally and transversely polarized vector $D$ (left panel) and $D_s$ (right panel) mesons as functions of the parameter $\beta$ (in GeV).} 
\label{fig:4}
\end{figure}

\begin{figure}[h]
\centering
\minipage{0.5\textwidth}
  \includegraphics[width=3 in]{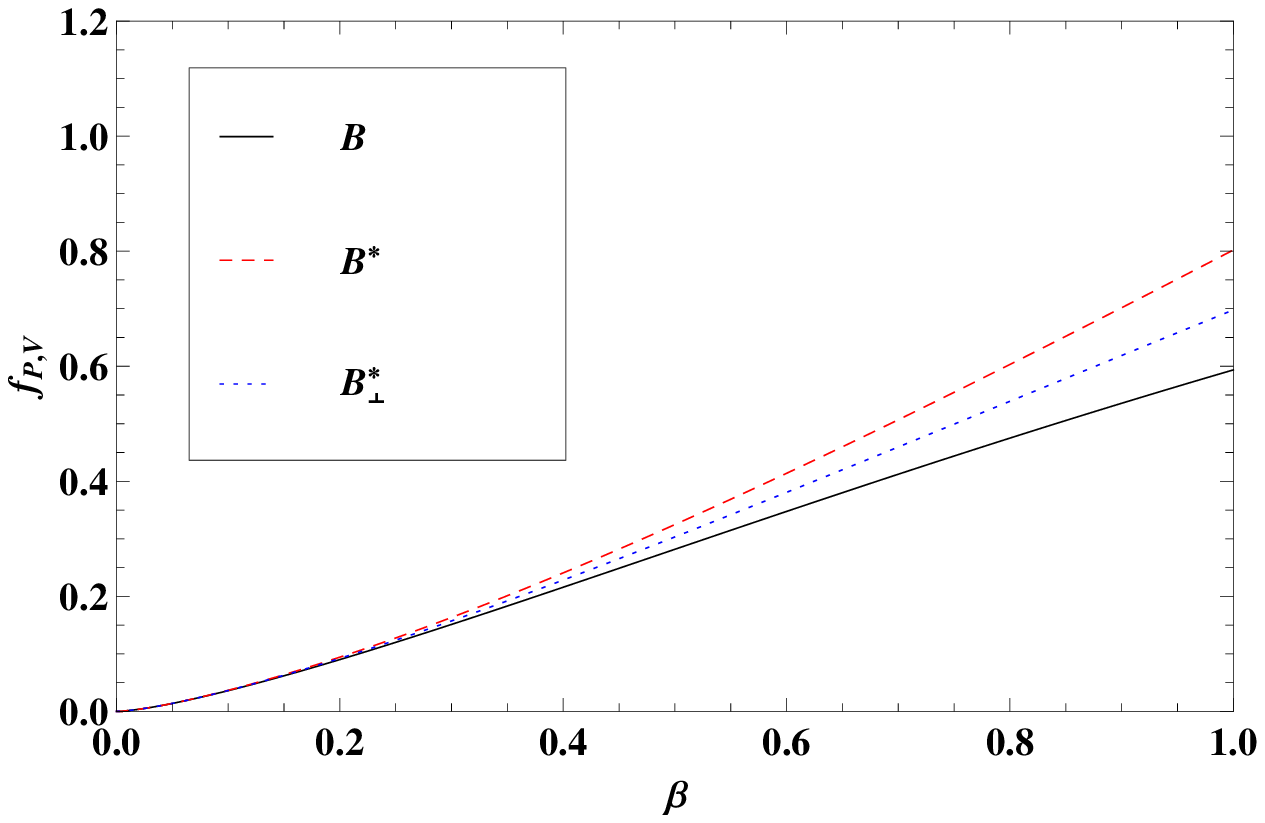}
\endminipage\hfill
\minipage{0.5\textwidth}
 \includegraphics[width=3 in]{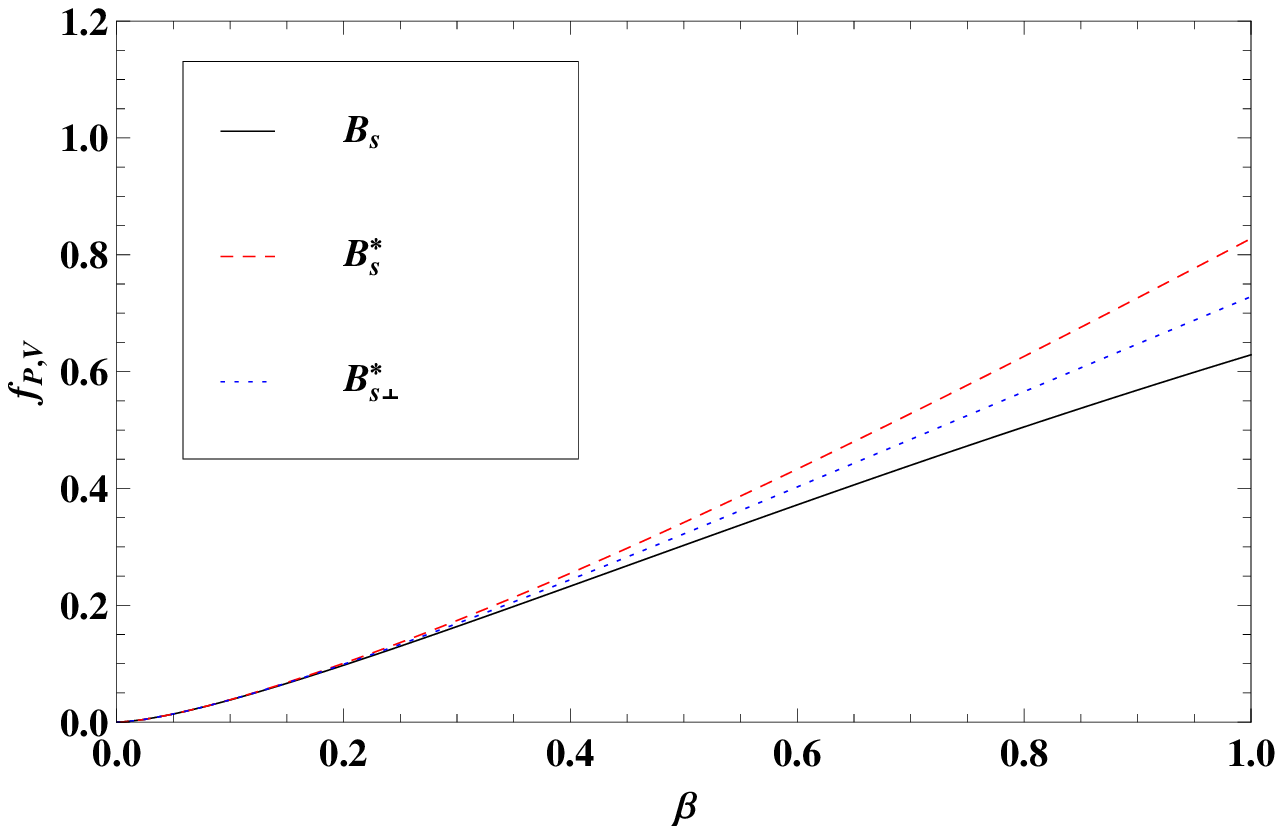}
\endminipage\hfill
\caption{The decay constants $f_{B_{(s)}}$, $f_{B^*_{(s)}}$ and $f_{B^*_{(s)}}^{\perp}$ of pseudoscalar, longitudinally and transversely polarized vector $B$ (left panel) and $B_s$ (right panel) mesons as functions of the parameter $\beta$ (in GeV).} 
\label{fig:5}
\end{figure}

In Fig. \ref{fig:2}, we have shown the dependence of decay constants of the non-strange light pseudoscalar ($\pi$) and the longitudinally and transversely polarized light vector mesons ($\rho$) on the parameter $\beta$. In Fig. \ref{fig:3}, we have presented the strange light pseudoscalar ($K$) and the longitudinally and transversely polarized strange light vector mesons ($K^*$) as a function of $\beta$. Comparing Figs.~\ref{fig:2} and~\ref{fig:3}, we find that the decay constants of non-strange light pseudoscalar and vector mesons are not much different from the decay constants of strange light pseudoscalar and vector mesons. In Fig. \ref{fig:4}, we have shown the dependence of decay constants of the heavy pseudoscalar ($D$ and $D_s$) and the longitudinally and transversely polarized heavy vector mesons ($D^*$ and $D_s^*$) on the parameter $\beta$. In Fig. \ref{fig:5}, we have shown the dependence of decay constants of the heavy pseudoscalar ($B$ and $B_s$) and the longitudinally and transversely polarized heavy vector mesons ($B^*$ and $B_s^*$) on $\beta$. As one may expect from the heavy quark symmetry, the difference between the heavy pseudoscalar mesons ($B$ and $B_s$) and the heavy vector mesons ($B^*$ and $B_s^*$) are substantially reduced in contrast to the difference between the light pseudoscalar mesons ($\pi$ and $K$) and the light vector mesons ($\rho$ and $K^*$). In general, the decay constants of the light and heavy mesons increase with the value of the parameter $\beta$. Especially, the longitudinally polarized vector meson decay constants are found to have the highest values followed by the transversely polarized vector mesons, and the pseudoscalar meson decay constants have the lowest values for given $\beta$ value.

\par In Table \ref{tab4}, we present our predictions for the decay constants of pseudoscalar, longitudinally and transversely polarized vector light ($\pi$, $\rho$, $K$, and $K^*$) mesons obtained using the parameters given in Tables \ref{tab1} and \ref{tab2} and compare them with the LFQM \cite{choi4, choi2}, BS model \cite{wang2}, RQM \cite{ebert}, SR \cite{ball0}, and lattice QCD \cite{khan1} predictions as well as with the available experimental data \cite{tana}. 
The first and second errors in our model calculations come from the $\pm 10\%$ variation of $\beta$ values for fixed quark masses 
and from the $\pm 10\%$ variation of quark masses for fixed $\beta$ values, respectively.
We note that our predictions for the pseudoscalar decay constants $f_\pi = 129^{+3.1+9.3}_{-3.7-9.8}$ MeV and $f_K = 158^{+6.2+8.8}_{-7.1-9.7}$ MeV are in good agreement with the available experimental data $f_\pi = 130.3\pm0.3$ MeV and $f_K = 156.1\pm0.5$ MeV \cite{tana}. 
Our model predictions are also consistent with the other theoretical model results. The values of ratios $f_{\rho}/f_{\pi}$, $f_{K^*}/f_{K}$, $f_{K}/f_{\pi}$, and $f_{K^*}/f_{\rho}$ for light mesons are listed in Table \ref{tab5} so as to have a deeper comprehension of the quantitative difference between pseudoscalar and vector mesons decay constants ($f_{V}/f_{P}$, $f_{P'}/f_{P}$, and $f_{V'}/f_{V}$). The comparison has also been made with the available experimental data \cite{tana} and other theoretical model calculations \cite{choi4, choi2, wang2, ebert, ball0, khan1}. It is evident that our results for the ratios 
$f_{K}/f_{\pi} = 1.22^{+0.018-0.019}_{-0.021+0.019}$ and $f_{K^*}/f_{\rho} = 1.04^{+0.004-0.005}_{-0.005+0.005}$ are not only comparable with the available experimental data \cite{tana} (
$f_{K}/f_{\pi} = 1.20\pm0.004$ and $f_{K^*}/f_{\rho} = 0.97\pm0.04$) 
but also consistent with the other theoretical model calculations.
In Table \ref{tab6}, we present our predictions for the decay constants of pseudoscalar, longitudinally and transversely polarized vector $D$ mesons and compare them with the LFQM \cite{choi1, choi2, hwang1, chao1}, LFHQCD \cite{dosch, chang1}, SR \cite{narison, wang1, gel}, lattice QCD \cite{aoki, bec, na2}, BS model \cite{wang2, cvetic, wang3}, RQM \cite{hwang2, capstick, ebert} and NRQM \cite{yazar2, yazar3, shady, hass} predictions as well as the available experimental data \cite{tana}. We note that our prediction for the decay constant 
$f_D = 197^{+19+0.2}_{-20-1.0}$ MeV
is comparable with the available experimental data ($f_D = 203.7\pm 4.7$ MeV) 
\cite{tana}. It is also observed that the theoretical results predicted in this work as well as in the other models differ from each other in one way or the other. One can see from Table \ref{tab6} that our predictions are in a reasonable agreement with the previous LFQM results, LFHQCD results, SR predictions and the lattice results. The different values of decay constants with respect to other theoretical models might be due to difference in the model assumptions or distinct choices of the model parameters. We have listed in Table \ref{tab7} our values of the ratios $f_{D^*}/f_{D}$, $f_{D_s^*}/f_{D_s}$, $f_{D_s}/f_{D}$ and $f_{D_s^*}/f_{D^*}$ and compared them with the available experimental data and other theoretical calculations. 
We can see that the ratios $f_{D_s}/f_{D} = 1.11^{-0.001-0.002}_{+0.001+0.002}$ and $f_{D_s^*}/f_{D^*} = 1.10^{-0.003-0.002}_{-0.002-0.003}$ in this work are in good agreement with $f_{D_s}/f_{D} = 1.11$ and $f_{D_s^*}/f_{D^*} = 1.13$ from the previous LFQM \cite{choi2}, $f_{D_s}/f_{D} = 1.09$ from LFHQCD \cite{dosch} and $f_{D_s}/f_{D} = 1.10 \pm 0.02$ and $f_{D_s^*}/f_{D^*} = 1.11 \pm 0.03$ from the lattice QCD \cite{bec}. Also from Table \ref{tab7}, we can observe that our results for the ratios $f_{D^*}/f_{D} = 1.17^{+0.03-0.03}_{-0.03+0.04}$ and $f_{D_s^*}/f_{D_s} = 1.16^{+0.03-0.03}_{-0.03+0.03}$ are also comparable with the other theoretical model calculations.
In Table \ref{tab8}, we present our predictions for the decay constants of pseudoscalar, longitudinally and transversely polarized vector $B$ mesons and compare them with the LFQM \cite{choi1, choi2, hwang1, chao1}, LFHQCD \cite{dosch, chang1}, SR \cite{narison, wang1, gel}, lattice QCD \cite{aoki, bec, na1}, BS model \cite{wang2, cvetic, wang3}, RQM \cite{hwang2, capstick, ebert} and NRQM \cite{yazar1, yazar3, shady, hass} predictions as well as with the available experimental data \cite{tana}. One can note that our prediction for the decay constant $f_B = 163^{+21-4}_{-20+4}$ MeV is quite comparable with the available experimental data $f_B = 188\pm 25$ MeV \cite{tana}. One can also observe that our model predictions are more or less comparable with the other theoretical model predictions. 
The difference in the values of decay constants with respect to other theoretical models might be because of the different model assumptions or distinct choices of the parameters. However, overall the predictions are fairly in the similar range. 
The values of the ratios $f_{B^*}/f_{B}$, $f_{B_s^*}/f_{B_s}$, $f_{B_s}/f_{B}$ and $f_{B_s^*}/f_{B^*}$ are also listed in Table \ref{tab9} and their comparison has been made with the other theoretical calculations as well. As one can see that our results for the ratios $f_{B_s}/f_{B} = 1.13^{-0.004+0.003}_{-0.003-0.003}$ and $f_{B_s^*}/f_{B^*}= 1.13^{-0.005+0.005}_{+0.006+0.001}$ are compatible with the previous LFQM results \cite{choi2} ( $f_{B_s}/f_{B} = 1.13$ and $f_{B_s^*}/f_{B^*} = 1.15$), SR predictions \cite{gel} ($f_{B_s}/f_{B} = 1.17^{+0.03}_{-0.04}$ and $f_{B_s^*}/f_{B^*} = 1.20\pm0.04$), the lattice results \cite{bec} ($f_{B_s}/f_{B} = 1.14\pm0.03^{+1}_{-1}$ and $f_{B_s^*}/f_{B^*} = 1.17\pm0.04^{+1}_{-3}$ ) 
and the RQM predictions \cite{ebert} ($f_{B_s}/f_{B} = 1.15$ and $f_{B_s^*}/f_{B^*} = 1.15$). One can also see that our results for the ratios $f_{B^*}/f_{B} = 1.06^{+0.010-0.011}_{-0.013+0.011}$ and $f_{B_s^*}/f_{B_s} = 1.05^{+0.008-0.010}_{-0.005+0.015}$ are comparable with the other theoretical model calculations.

\par We should note that in the case of heavy mesons, the ratios $f_V/f_P$ of the vector and the pseudoscalar mesons are larger for the case of $D_{(s)}$ mesons than for the case of $B_{(s)}$ mesons as one may expect from the heavy quark symmetry. This may also be accounted from the last terms of Eqs. (\ref{eqn:24}) and (\ref{eqn:25}) proportional to $\frac{\textbf{k}_{\perp}^2}{M_0 + m_q + m_{\bar q}}$. As ${\langle\textbf{k}_{\perp}^2\rangle}_{q \bar q}^{1/2} = {\beta}_{q \bar q}$ at least for 1$S$ basis, the quark mass dominates over the scale parameter $\beta$ leading to the ratio $f_V/f_P$ for $D$ mesons being larger than that of $B$ mesons.
Our results are also consistent with the model-independent analysis of semileptonic $B$ meson decays in the context of the heavy quark effective theory \cite{neubert1}.

\begin{figure}[h]
\centering
  \includegraphics[width=3.5 in]{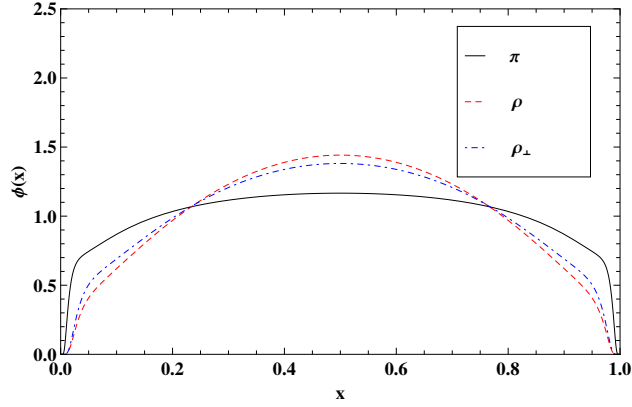}
\caption{Normalized DAs for pseudoscalar $\pi$ (solid line), longitudinally polarized vector $\rho$ (dashed line) and transversely 
polarized vector $\rho$ (dotted dashed line) mesons.} 
\label{fig:6}
\end{figure}

\begin{figure}[h]
\centering
 \includegraphics[width=3.5 in]{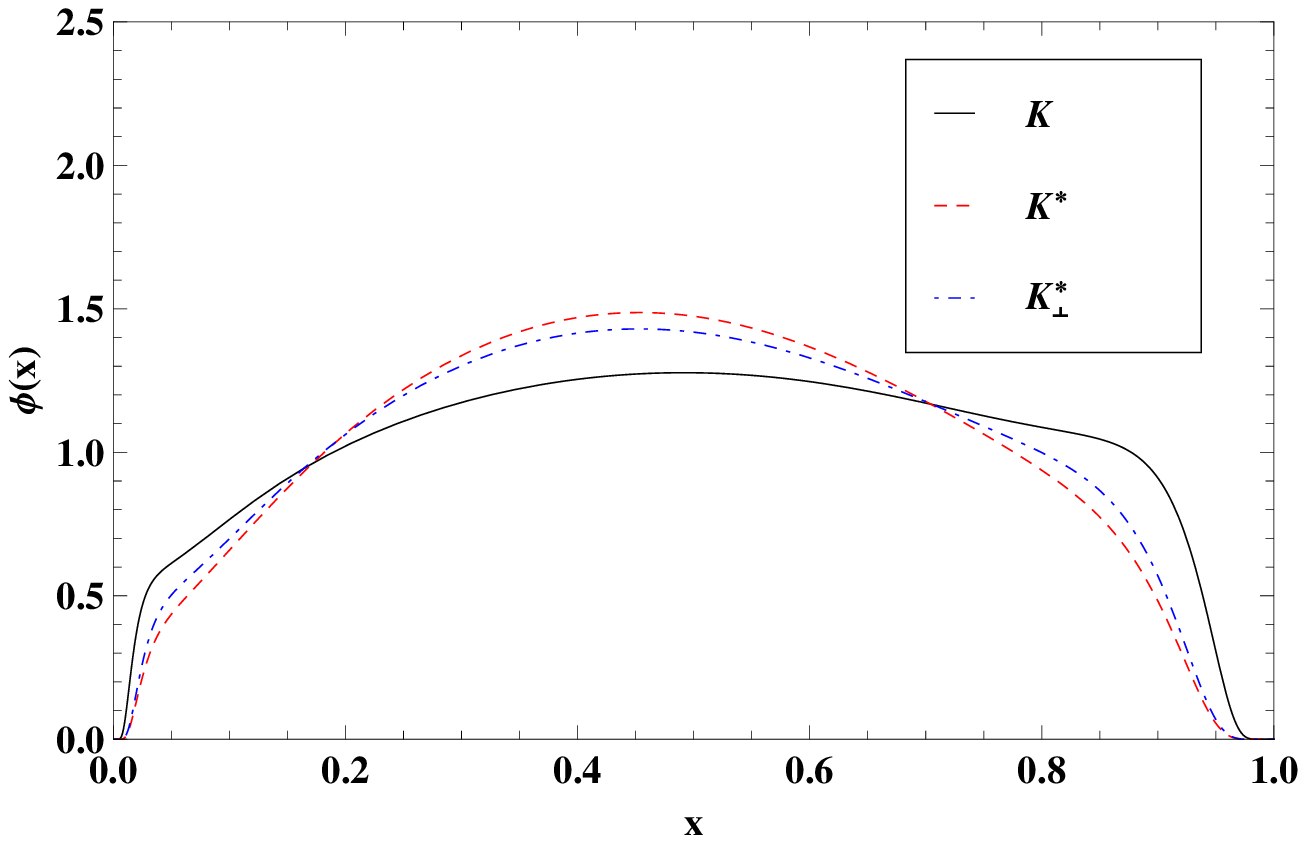}
\caption{Normalized DAs for the pseudoscalar $K$ (solid line), longitudinally polarized vector $K$ (dashed line) and transversely polarized vector $K$ (dotted dashed line) mesons.} 
\label{fig:7}
\end{figure}

\begin{figure}[h]
\centering
\minipage{0.5\textwidth}
  \includegraphics[width=3 in]{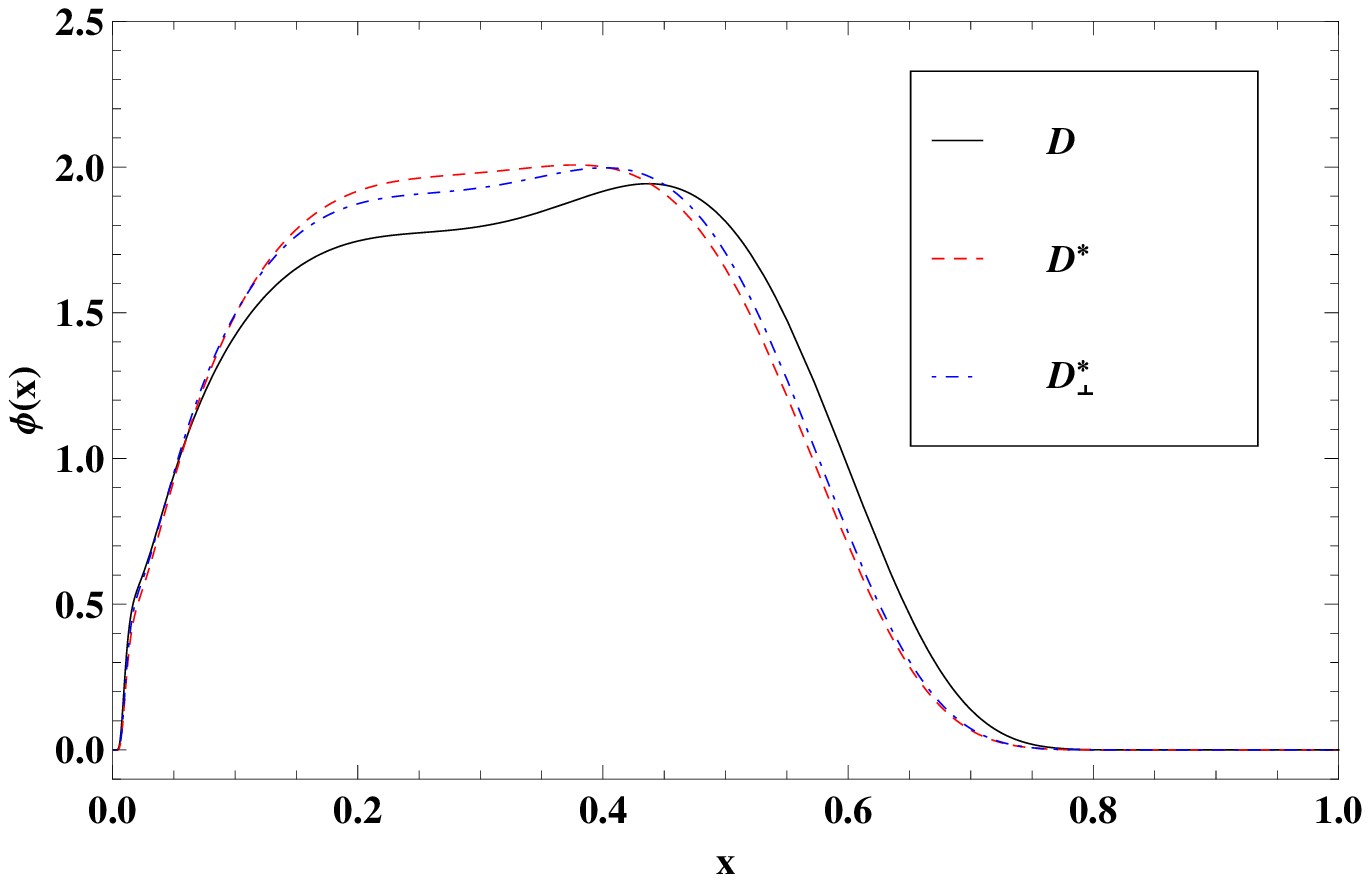}
\endminipage\hfill
\minipage{0.5\textwidth}
 \includegraphics[width=3 in]{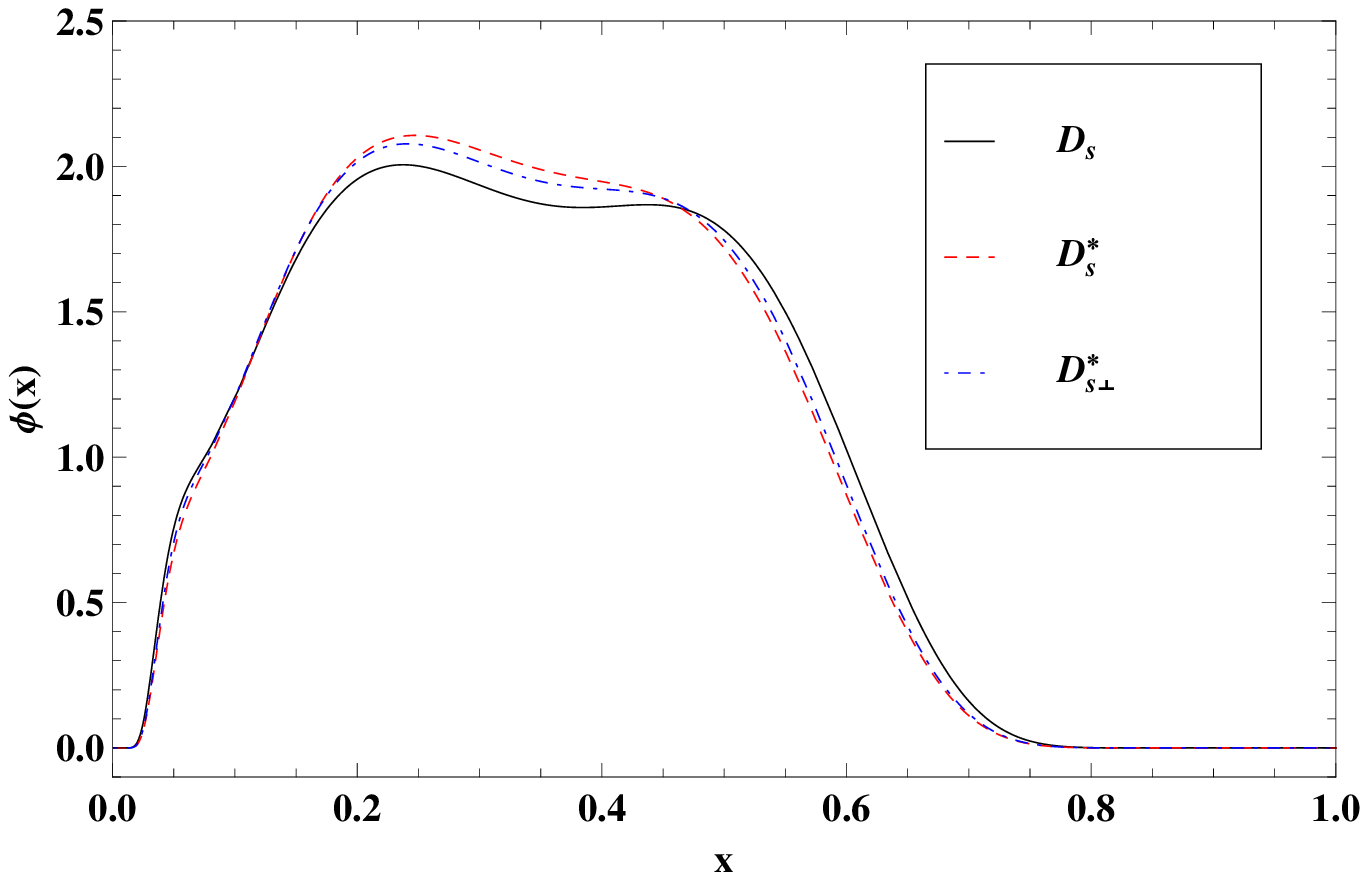}
\endminipage\hfill
\caption{Normalized DAs for the heavy pseudoscalar (solid line), longitudinally (dashed line) and transversely (dotted dashed line) polarized vector $D$ (left panel) and $D_s$ (right panel) mesons.} 
\label{fig:8}
\end{figure}

\begin{figure}[h]
\centering
\minipage{0.5\textwidth}
  \includegraphics[width=3 in]{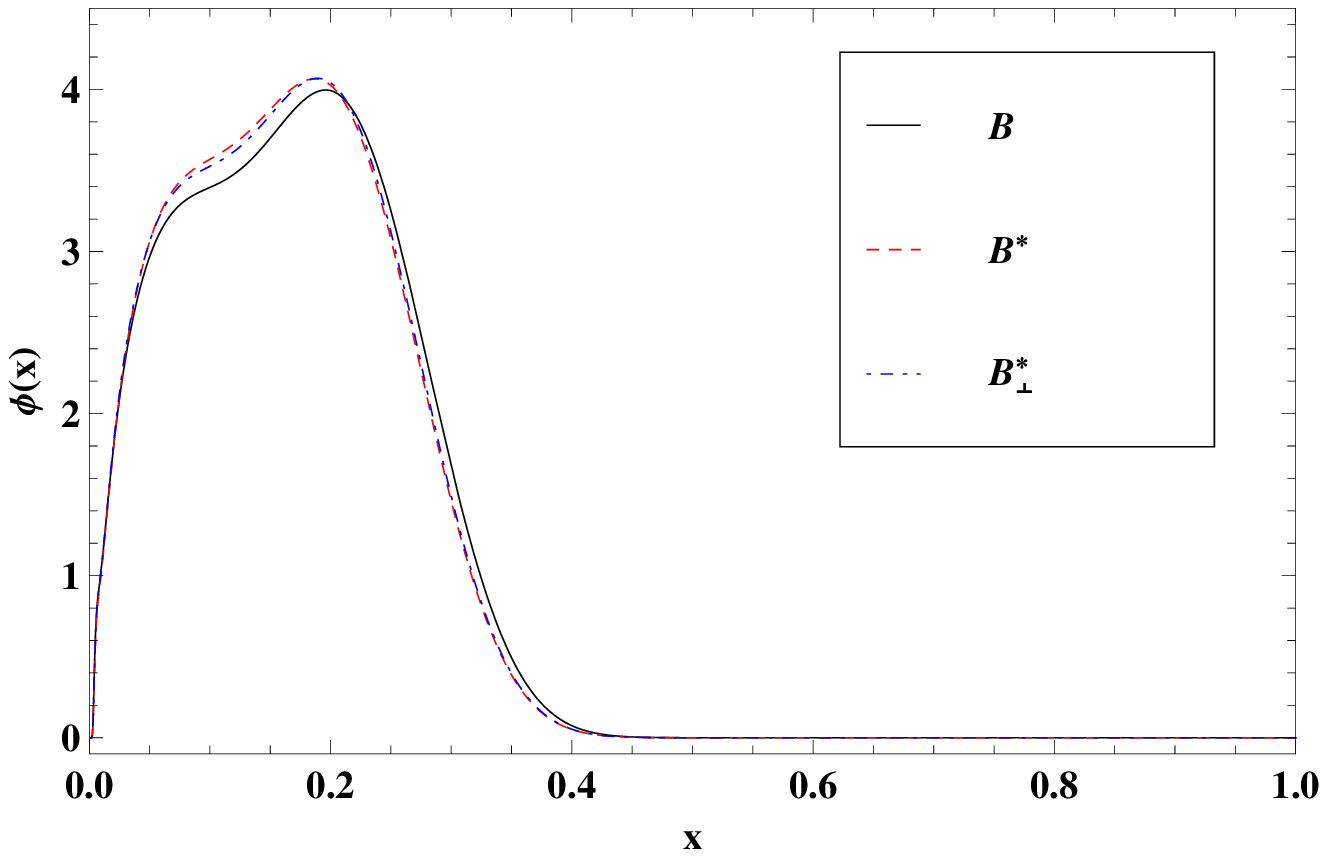}
\endminipage\hfill
\minipage{0.5\textwidth}
 \includegraphics[width=3 in]{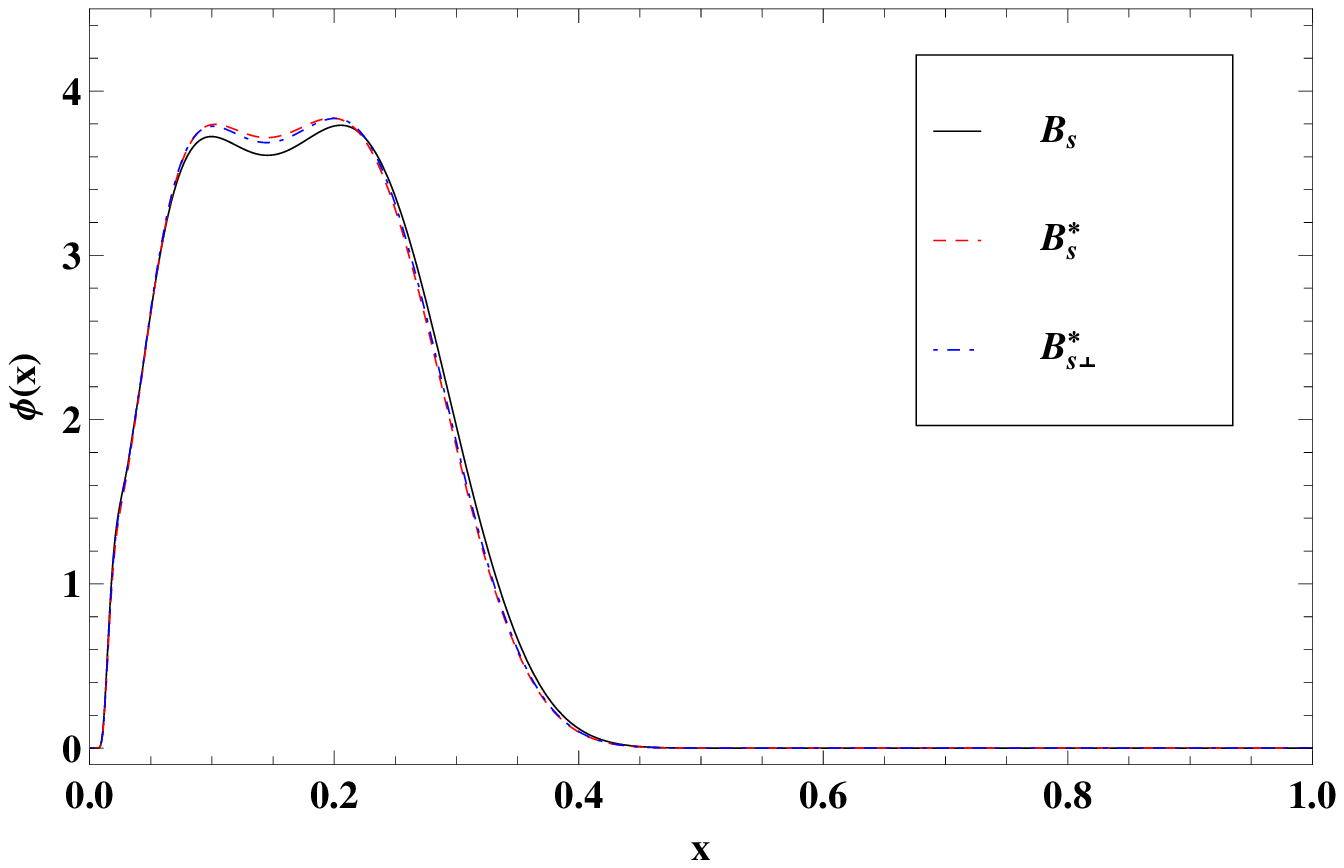}
\endminipage\hfill
\caption{Normalized DAs for the heavy pseudoscalar (solid line), longitudinally (dashed line) and transversely (dotted dashed line) polarized vector $B$ (left panel) and $B_s$ (right panel) mesons.} 
\label{fig:9}
\end{figure}

We show in Figs. \ref{fig:6}$-$\ref{fig:9} the normalized quark DAs ($\phi_P (x)$, $\phi_{V\parallel} (x)$ and $\phi_{V\perp} (x)$) for pseudoscalar (solid line), longitudinally polarized vector (dashed line) and transversely polarized vector (dotted dashed line) light ($\pi$, $\rho$, $K$, and $K^*$) and heavy ($D$, $D^*$, $D_s$, $D_s^*$ and $B$, $B^*$, $B_s$ and $B_s^*$) mesons. Since the quark mass and the parameter $\beta$ are same for the $\phi_{V\parallel} (x)$ and $\phi_{V\perp} (x)$ DAs, the difference between them is very small. Due to the flavor SU(3)-symmetry breaking effect, the quark DAs of $K$ mesons show the asymmetric feature in comparison with that of $\pi$ mesons. In the case of heavy mesons, the quark DAs' peaks of $B$, $B^*$, $B_s$ and $B_s^*$ mesons are much narrower than those of $D$, $D^*$, $D_s$ and $D_s^*$ mesons due to the large mass difference between $b$ and $c$ quarks. The strange quark effect appears relatively more pronounced in $D_s$ and $D_s^*$ than in $B_s$ and $B_s^*$, which may also be understood from the heavy quark symmetry.
To further exploit such relative effects, we present in Table \ref{tab10} and \ref{tab11} the first six $\xi$-moments for $D$ and $B$ mesons, respectively. It is found from the tables that the $\xi$-moments of $B$ mesons are higher in magnitude as compared to the $D$ mesons.
 
\section{Summary and Conclusions}
\label{sec5}
In this work, we have studied the mass spectra, decay constants and the twist-2 DAs of pseudoscalar and vector light ($\pi$, $K$) and heavy ($D$, $D_s$, $B$, $B_s$) mesons within the LFQM with the exponential-type confining potential. Our LFQM is constrained by the variational principle for the QCD-motivated effective Hamiltonian not only with the confining potential but also with the Coulomb and hyperfine interaction. We have used a Gaussian smearing function to weaken the singularity of the Dirac $\delta$ function in hyperfine interaction so as to avoid the issue of negative infinity~\cite{choi2}. 
We have calculated the mass spectra of the ground state pseudoscalar and vector light and heavy mesons as well as the decay constants of the corresponding mesons using the mixed wave function $\Phi$ of $1S$, $2S$ and $3S$ HO states as the trial wave function. We also compared our results with available experimental data and the previous LFQM results with the linear and HO potentials~\cite{choi1,choi2} as well as other theoretical model predictions. We note that our LFQM predictions are comparable to each other regardless of the confining potential type as far as the efficacy of model prediction is achieved by allowing sufficient number of HO basis functions for the trial wave function. It appears however that we need more HO basis functions for the trial wave function in the case of the exponential-type confining potential compare to the case of the linear and HO confining potentials. 
For the present analysis with the exponential-type confining potential, we used the larger number of HO basis functions ($1S$, $2S$, and $3S$) to achieve the efficacy of the model calculations, while we achieved the efficacy of LFQM with the linear and HO confining potentials using only up to the two lowest order HO wave functions in our previous analyses~\cite{choi1,choi2}.

Not only for the mass sprectra but also for the decay constants of pseudoscalar and vector light and heavy mesons, our results are in reasonable agreement with the available experimental data as well as comparable with other theoretical model predictions. Our results for the quark DAs of $K$ mesons show the asymmetric feature in comparison with that of $\pi$ mesons due to the flavor SU(3)-symmetry breaking effect. Consistent with the heavy quark symmetry, our results for the ratios $f_V/f_P$ of the vector and the pseudoscalar $D$ mesons are larger in comparison with the $B$ mesons. Also, the quark DAs of $D$, $D^*$, $D_s$ and $D^*_s$ mesons show much broader shapes than those of $B$, $B^*$, $B_s$ and $B^*_s$ mesons due to the large mass difference between $b$ and $c$ quarks.
\par Even though the observables that we have estimated have been previously calculated in the light-front framework, the present work is based on the idea of modelling the potential in a rather significantly different way from the previous works~\cite{choi1,choi2}. We anticipate to study further other wave function related observables such as meson transition form factors and also analyse radially excited meson states using the larger number of HO basis functions.

\acknowledgements
N.D. would like to thank Monika Randhawa (University Institute of Engineering and
Technology, Panjab University) for helpful discussions. H.D. would like to thank the Department of Science and Technology (Ref No. EMR/2017/001549) Government of India for financial support. C.-R. Ji was supported in part by the US Department of Energy (Grant No. DE-FG02-03ER41260).
H.-M. Choi was supported by the National Research Foundation of Korea (NRF) (Grant No. NRF-2017R1D1A1B03033129).

\newpage 
\begin{table}[h]
\centering
\caption{Decay constants (in units of MeV) for light mesons ($\pi$, $\rho$, $K$ and $K^*$) in the present work and their comparison with the available experimental data and other theoretical model predictions. The first and second errors  in the present work come from the $\pm 10\%$ variation of $\beta$ values  fixed quark masses and from the $\pm 10\%$ variation of quark masses  for fixed $\beta$ values, respectively.}
\vspace{0.2 cm}
\label{tab4}
\begin{tabular}{lcccccc}
\hline\hline\\[-1.5 ex]
\multicolumn{1}{c}{} &\hspace{0.1in}{$f_{\pi}$} &\hspace{0.1in}{$f_{\rho}$} &\hspace{0.1in}{$f_{\rho}^{\perp}$} &\hspace{0.1in}{$f_{K}$} &\hspace{0.1in}{$f_{K^*}$} &\hspace{0.1in}{$f_{K^*}^{\perp}$} \\[1 ex]
\hline\\[-1.5 ex]

{Present work} &\hspace{0.0in}$129^{+3.1+9.3}_{-3.7-9.8}$ &\hspace{0.0in}$242^{+23.3+0.7}_{-23.5-0.9}$ &\hspace{0.0in}$178^{+13.5+3.9}_{-13.9-4.3}$
&\hspace{0.0in}$158^{+6.2+8.8}_{-7.1-9.7}$ 
&\hspace{0.0in}$253^{+25.5-0.5}_{-25.8+0.2}$ &\hspace{0.0in}$199^{+16.1+3.3}_{-16.7-3.9}$ \\[1 ex]

{Exp. \cite{tana}} &\hspace{0.2in}{$130.3\pm0.3$} &\hspace{0.2in}{$210\pm4$} &\hspace{0.2in}{$-$} &\hspace{0.2in}{$156.1\pm0.5$} &\hspace{0.2in}{$204\pm7$} &\hspace{0.2in}{$-$} \\[1 ex]

{LFQM, Lin \cite{choi4}} &\hspace{0.2in}{130} &\hspace{0.2in}{246} &\hspace{0.2in}{188} &\hspace{0.2in}{161} &\hspace{0.2in}{256} &\hspace{0.2in}{210} \\[1 ex]

{LFQM, HO \cite{choi4}} &\hspace{0.2in}{131} &\hspace{0.2in}{215} &\hspace{0.2in}{173} &\hspace{0.2in}{155} &\hspace{0.2in}{223} &\hspace{0.2in}{191} \\[1 ex]

{LFQM, Lin \cite{choi2}} &\hspace{0.2in}{130} &\hspace{0.2in}{205} &\hspace{0.2in}{$-$} &\hspace{0.2in}{161} &\hspace{0.2in}{224} &\hspace{0.2in}{$-$} \\[1 ex]

{BS \cite{wang2}} &\hspace{0.2in}{127} &\hspace{0.2in}{$-$} &\hspace{0.2in}{$-$} &\hspace{0.2in}{157} &\hspace{0.2in}{$-$} &\hspace{0.2in}{$-$} \\[1 ex]

{RQM \cite{ebert}} &\hspace{0.2in}{124} &\hspace{0.2in}{219} &\hspace{0.2in}{$-$} &\hspace{0.2in}{155} &\hspace{0.2in}{236} &\hspace{0.2in}{$-$} \\[1 ex]

{SR \cite{ball0}} &\hspace{0.2in}{$-$} &\hspace{0.2in}{$205\pm9$} &\hspace{0.2in}{$160\pm10$} &\hspace{0.2in}{$-$} &\hspace{0.2in}{$217\pm5$} &\hspace{0.2in}{$170\pm10$} \\[1 ex]

{Lattice QCD \cite{khan1}} &\hspace{0.2in}{$126.6\pm6.4$} &\hspace{0.2in}{$239.4\pm7.3$} &\hspace{0.2in}{$-$} &\hspace{0.2in}{$152.0\pm6.1$} &\hspace{0.2in}{$255.5\pm6.5$} &\hspace{0.2in}{$-$} \\[1 ex]

\hline\hline
\end{tabular}
\end{table}

\begin{table}[h]
\centering
\caption{Ratio of the decay constants for light mesons ($\pi$, $\rho$, $K$ and $K^*$) compared with the available experimental data and other theoretical model calculations.}
\vspace{0.2 cm}
\label{tab5}
\begin{tabular}{lcccc}
\hline\hline\\[-1.5 ex]
\multicolumn{1}{c}{} &\hspace{0.6in}{$f_{\rho}/f_{\pi}$} & \hspace{0.5in}{$f_{K^*}/f_{K}$} &\hspace{0.5in}{$f_{K}/f_{\pi}$} &\hspace{0.5in}{$f_{K^*}/f_{\rho}$} \\[1 ex]
\hline\\[-1.5 ex]

{Present work} &\hspace{0.3in}$1.88^{+0.13-0.12}_{-0.14+0.15}$ &\hspace{0.3in}$1.60^{+0.10-0.09}_{-0.09+0.10}$ &\hspace{0.3in}$1.22^{+0.018-0.019}_{-0.021+0.019}$
&\hspace{0.3in}$1.04^{+0.004-0.005}_{-0.005+0.005}$ \\[1 ex]

{Exp. \cite{tana}} &\hspace{0.6in}{$1.61\pm0.03$} &\hspace{0.5in}{$1.31\pm0.04$} &\hspace{0.5in}{$1.20\pm0.004$} &\hspace{0.5in}{$0.97\pm0.04$} \\[1 ex]

{LFQM, Lin \cite{choi4}} &\hspace{0.6in}{1.89} &\hspace{0.5in}{1.59} &\hspace{0.5in}{1.24} &\hspace{0.5in}{1.04} \\[1 ex]

{LFQM, HO \cite{choi4}} &\hspace{0.6in}{1.64} &\hspace{0.5in}{1.44} &\hspace{0.5in}{1.18} &\hspace{0.5in}{1.04} \\[1 ex]

{LFQM, Lin \cite{choi2}} &\hspace{0.6in}{1.58} &\hspace{0.5in}{1.39} &\hspace{0.5in}{1.24} &\hspace{0.5in}{1.09} \\[1 ex]

{BS \cite{ebert}} &\hspace{0.6in}{$-$} &\hspace{0.5in}{$-$} &\hspace{0.5in}{1.24} &\hspace{0.5in}{$-$} \\[1 ex]

{SR \cite{ball0}} &\hspace{0.6in}{$-$} &\hspace{0.5in}{$-$} &\hspace{0.5in}{$-$} &\hspace{0.5in}{$1.06\pm0.05$} \\[1 ex]

{Lattice QCD \cite{khan1}} &\hspace{0.6in}{$1.90\pm0.11$} &\hspace{0.5in}{$1.68\pm0.08$} &\hspace{0.5in}{$1.20\pm0.08$} &\hspace{0.5in}{$1.07\pm0.04$} \\[1 ex]

\hline\hline
\end{tabular}
\end{table}

\begin{table}[h]
\centering
\caption{
Pseudoscalar, longitudinally and transversely polarized vector $D$ meson decay constants (in units of MeV) in the present work and their comparison with the available experimental data and other theoretical model predictions.}
\vspace{0.2 cm}
\label{tab6}
\begin{tabular}{p{1.35 in}m{0.85 in}p{0.80 in}p{0.80 in}p{0.85 in}p{0.80 in}p{0.75 in}}
\hline\hline\\[-1.5 ex]
\multicolumn{1}{c}{} &\multicolumn{1}{c}{$f_{D}$} & \multicolumn{1}{c}{$f_{D^*}$} &\multicolumn{1}{c}{$f_{D^*}^{\perp}$} &\multicolumn{1}{c}{$f_{D_s}$} &\multicolumn{1}{c}{$f_{D_s^*}$} &\multicolumn{1}{c}{$f_{D_s^*}^{\perp}$} \\[1 ex]
\hline\\[-1.5 ex]

{Present work} &\multicolumn{1}{c}{$197^{+19+0.2}_{-20-1.0}$} &\multicolumn{1}{c}{$230^{+29-5}_{-28+6}$} &\multicolumn{1}{c}{$208^{+24-3}_{-24+3}$} 
&\multicolumn{1}{c}{$219^{+21-0.2}_{-22-0.8}$} 
&\multicolumn{1}{c}{$253^{+31-6}_{-31+6}$} &\multicolumn{1}{c}{$233^{+26-3}_{-26+3}$}\\[1 ex]

{Exp. \cite{tana}} &\multicolumn{1}{c}{$203.7\pm 4.7$} &\quad \quad{$-$} &\quad \quad{$-$} &\multicolumn{1}{c}{$257.8\pm 4.1$} &\quad \quad{$-$} &\quad \quad{$-$}\\[1 ex]

{LFQM, Lin \cite{choi1}} &\multicolumn{1}{c}{197} &\multicolumn{1}{c}{239} &\multicolumn{1}{c}{$-$} &\multicolumn{1}{c}{233} &\multicolumn{1}{c}{274} &\multicolumn{1}{c}{$-$}\\[1 ex]

{LFQM, HO \cite{choi1}} &\multicolumn{1}{c}{180} &\multicolumn{1}{c}{212} &\multicolumn{1}{c}{$-$} &\multicolumn{1}{c}{218} &\multicolumn{1}{c}{252} &\multicolumn{1}{c}{$-$}\\[1 ex]

{LFQM, Lin \cite{choi2}} &\multicolumn{1}{c}{208} &\multicolumn{1}{c}{230} &\multicolumn{1}{c}{$-$} &\multicolumn{1}{c}{231} &\multicolumn{1}{c}{260} &\multicolumn{1}{c}{$-$}\\[1 ex]

\multirow{2}{5em}{LFQM \cite{hwang1}} &\multicolumn{1}{c}{$-$} &\multicolumn{1}{c}{$259.6\pm14.6$} &\multicolumn{1}{c}{$232.7\pm11.7$} &\multicolumn{1}{c}{$267.4\pm17.9$} &\multicolumn{1}{c}{$338.7\pm29.7$} &\multicolumn{1}{c}{$303.1\pm23.8$} \\
&\multicolumn{1}{c}{} &\multicolumn{1}{c}{$306.3^{+18.2}_{-17.7}$} &\multicolumn{1}{c}{$256.2^{+13.6}_{-13.3}$} &\multicolumn{1}{c}{$259.7\pm13.7$} &\multicolumn{1}{c}{$391\pm28.9$} &\multicolumn{1}{c}{$325.3\pm21.5$} \\[1 ex]

{LFQM \cite{chao1}} &\multicolumn{1}{c}{$-$} &\multicolumn{1}{c}{$252.0^{+13.8}_{-11.6}$} &\multicolumn{1}{c}{$-$} &\multicolumn{1}{c}{$-$} &\multicolumn{1}{c}{$318.3^{+15.3}_{-12.6}$} &\multicolumn{1}{c}{$-$}\\[1 ex]

{LFHQCD \cite{dosch}} &\multicolumn{1}{c}{$199$} &\multicolumn{1}{c}{$-$} &\multicolumn{1}{c}{$-$} &\multicolumn{1}{c}{$216$} &\multicolumn{1}{c}{$-$} &\multicolumn{1}{c}{$-$}\\[1 ex]

{LFHQCD \cite{chang1}} &\multicolumn{1}{c}{$214.2^{+7.6}_{-7.8}$} &\multicolumn{1}{c}{$-$} &\multicolumn{1}{c}{$-$} &\multicolumn{1}{c}{$253.5^{+6.6}_{-7.1}$} &\multicolumn{1}{c}{$-$} &\multicolumn{1}{c}{$-$}\\[1 ex]

{SR \cite{narison}} &\multicolumn{1}{c}{$204\pm4.6$} &\multicolumn{1}{c}{$-$} &\multicolumn{1}{c}{$-$} &\multicolumn{1}{c}{$243.2\pm4.9$} &\multicolumn{1}{c}{$-$} &\multicolumn{1}{c}{$-$}\\[1 ex]

{SR \cite{wang1}} &\multicolumn{1}{c}{$208\pm10$} &\multicolumn{1}{c}{$263\pm21$} &\multicolumn{1}{c}{$-$} &\multicolumn{1}{c}{$240\pm10$} &\multicolumn{1}{c}{$308\pm21$} &\multicolumn{1}{c}{$-$}\\[1 ex]

{SR \cite{gel}} &\multicolumn{1}{c}{$201^{+12}_{-13}$} &\multicolumn{1}{c}{$242^{+20}_{-12}$} &\multicolumn{1}{c}{$-$} &\multicolumn{1}{c}{$238^{+13}_{-23}$} &\multicolumn{1}{c}{$293^{+19}_{-14}$} &\multicolumn{1}{c}{$-$}\\[1 ex]

{Lattice QCD \cite{aoki}} &\multicolumn{1}{c}{$211.9\pm1.1$} &\multicolumn{1}{c}{$-$} &\multicolumn{1}{c}{$-$} &\multicolumn{1}{c}{$249\pm1.2$} &\multicolumn{1}{c}{$-$} &\multicolumn{1}{c}{$-$}\\[1 ex]

{Lattice QCD \cite{bec}} &\multicolumn{1}{c}{$211\pm14^{+2}_{-12}$} &\multicolumn{1}{c}{$245\pm20^{+3}_{-2}$} &\multicolumn{1}{c}{$-$} &\multicolumn{1}{c}{$231\pm12^{+8}_{-1}$} &\multicolumn{1}{c}{$272\pm16^{+3}_{-20}$} &\multicolumn{1}{c}{$-$}\\[1 ex]

{Lattice QCD \cite{na2}} &\multicolumn{1}{c}{$-$} &\multicolumn{1}{c}{$-$} &\multicolumn{1}{c}{$-$} &\multicolumn{1}{c}{$248\pm25$} &\multicolumn{1}{c}{$-$} &\multicolumn{1}{c}{$-$}\\[1 ex]

{BS \cite{wang2}} &\multicolumn{1}{c}{$238$} &\multicolumn{1}{c}{$-$} &\multicolumn{1}{c}{$-$} &\multicolumn{1}{c}{$241$} &\multicolumn{1}{c}{$-$} &\multicolumn{1}{c}{$-$}\\[1 ex]

{BS \cite{cvetic, wang3}} &\multicolumn{1}{c}{$230\pm25$} &\multicolumn{1}{c}{$340\pm23$} &\multicolumn{1}{c}{$-$} &\multicolumn{1}{c}{$248\pm27$} &\multicolumn{1}{c}{$375\pm24$} &\multicolumn{1}{c}{$-$}\\[1 ex]

{RQM \cite{hwang2}} &\multicolumn{1}{c}{$271\pm14$} &\multicolumn{1}{c}{$327\pm13$} &\multicolumn{1}{c}{$-$} &\multicolumn{1}{c}{$309\pm15$} &\multicolumn{1}{c}{$362\pm15$} &\multicolumn{1}{c}{$-$}\\[1 ex]

{RQM \cite{capstick}} &\multicolumn{1}{c}{$240\pm20$} &\multicolumn{1}{c}{$-$} &\multicolumn{1}{c}{$-$} &\multicolumn{1}{c}{$290\pm20$} &\multicolumn{1}{c}{$-$} &\multicolumn{1}{c}{$-$}\\[1 ex]

{RQM \cite{ebert}} &\multicolumn{1}{c}{$234$} &\multicolumn{1}{c}{$310$} &\multicolumn{1}{c}{$-$} &\multicolumn{1}{c}{$268$} &\multicolumn{1}{c}{$315$} &\multicolumn{1}{c}{$-$}\\[1 ex]

{NRQM \cite{yazar2}} &\multicolumn{1}{c}{$318$} &\multicolumn{1}{c}{$307$} &\multicolumn{1}{c}{$-$} &\multicolumn{1}{c}{$354$} &\multicolumn{1}{c}{$344$} &\multicolumn{1}{c}{$-$}\\[1 ex]

{NRQM \cite{yazar3}} &\multicolumn{1}{c}{$368.8$} &\multicolumn{1}{c}{$353.8$} &\multicolumn{1}{c}{$-$} &\multicolumn{1}{c}{$394.8$} &\multicolumn{1}{c}{$382.1$} &\multicolumn{1}{c}{$-$}\\[1 ex]

{NRQM \cite{shady}} &\multicolumn{1}{c}{$220$} &\multicolumn{1}{c}{$290$} &\multicolumn{1}{c}{$-$} &\multicolumn{1}{c}{$250$} &\multicolumn{1}{c}{$310$} &\multicolumn{1}{c}{$-$}\\[1 ex]

{NRQM \cite{hass}} &\multicolumn{1}{c}{$228$} &\multicolumn{1}{c}{$-$} &\multicolumn{1}{c}{$-$} &\multicolumn{1}{c}{$273$} &\multicolumn{1}{c}{$-$} &\multicolumn{1}{c}{$-$}\\[1 ex]

\hline\hline
\end{tabular}
\end{table}

\begin{table}[h]
\centering
\caption{Ratio of the decay constants for $(D, D_s, D^*, D^*_s)$ mesons compared with the available experimental data and other theoretical model calculations.}
\vspace{0.2 cm}
\label{tab7}
\begin{tabular}{lcccc}
\hline\hline\\[-1.5 ex]
\multicolumn{1}{c}{} &\hspace{0.6in}{$f_{D^*}/f_{D}$} & \hspace{0.5in}{$f_{D_s^*}/f_{D_s}$} &\hspace{0.5in}{$f_{D_s}/f_{D}$} &\hspace{0.5in}{$f_{D_s^*}/f_{D^*}$} \\[1 ex]
\hline\\[-1.5 ex]

{Present work} &\hspace{0.in}{$1.17^{+0.03-0.03}_{-0.03+0.04}$} &\hspace{0.in}{$1.16^{+0.03-0.03}_{-0.03+0.03}$} &\hspace{0.in}{$1.11^{-0.001-0.002}_{+0.001+0.002}$} 
&\hspace{0.in}{$1.10^{-0.003-0.002}_{-0.001-0.003}$}\\[1 ex]

{Exp. \cite{tana}} &\hspace{0.6in}{$-$} &\hspace{0.5in}{$-$} &\hspace{0.5in}{$1.27\pm 0.03$} &\hspace{0.5in}{$-$}\\[1 ex]

{LFQM, Lin \cite{choi1}} &\hspace{0.6in}{1.21} &\hspace{0.5in}{1.18} &\hspace{0.5in}{1.18} &\hspace{0.5in}{1.15}\\[1 ex]

{LFQM, HO \cite{choi1}} &\hspace{0.6in}{1.18} &\hspace{0.5in}{1.16} &\hspace{0.5in}{1.21} &\hspace{0.5in}{1.19}\\[1 ex]

{LFQM, Lin \cite{choi2}} &\hspace{0.6in}{1.11} &\hspace{0.5in}{1.13} &\hspace{0.5in}{1.11} &\hspace{0.5in}{1.13}\\[1 ex]

\multirow{2}{5em}{LFQM \cite{hwang1}} &\hspace{0.6in}{$1.26\pm0.02$} &\hspace{0.5in}{$1.27\pm0.03$} &\hspace{0.5in}{$1.30\pm0.04$} &\hspace{0.5in}{$1.30\pm0.05$} \\
&\hspace{0.6in}{$1.49\pm0.02$} &\hspace{0.5in}{$1.51\pm0.03$} &\hspace{0.5in}{$1.26\pm0.04$} &\hspace{0.5in}{$1.28\pm0.05$} \\[1 ex]

{LFQM \cite{chao1}} &\hspace{0.6in}{$1.232^{+0.074}_{-0.064}$} &\hspace{0.5in}{$1.236^{+0.063}_{-0.054}$} &\hspace{0.5in}{$-$} &\hspace{0.5in}{$-$} \\[1 ex]

{LFHQCD \cite{dosch}} &\hspace{0.6in}{$-$} &\hspace{0.5in}{$-$} &\hspace{0.5in}{$1.09$} &\hspace{0.5in}{$-$} \\[1 ex]

{LFHQCD \cite{chang1}} &\hspace{0.6in}{$-$} &\hspace{0.5in}{$-$} &\hspace{0.5in}{$1.184^{+0.054}_{-0.052}$} &\hspace{0.5in}{$-$} \\[1 ex]

{SR \cite{narison}} &\hspace{0.6in}{$1.215\pm30$} &\hspace{0.5in}{$-$} &\hspace{0.5in}{$1.170\pm23$} &\hspace{0.5in}{$1.16\pm4$} \\[1 ex]

{SR \cite{wang1}} &\hspace{0.6in}{$-$} &\hspace{0.5in}{$-$} &\hspace{0.5in}{$1.15\pm0.06$} &\hspace{0.5in}{$-$} \\[1 ex]

{SR \cite{gel}} &\hspace{0.6in}{$1.20^{+0.13}_{-0.07}$} &\hspace{0.5in}{$1.24^{+0.13}_{-0.05}$} &\hspace{0.5in}{$1.18^{+0.04}_{-0.05}$} &\hspace{0.5in}{$1.21\pm0.05$} \\[1 ex]

{Lattice QCD \cite{aoki}} &\hspace{0.6in}{$-$} &\hspace{0.5in}{$-$} &\hspace{0.5in}{$1.173\pm0.003$} &\hspace{0.5in}{$-$} \\[1 ex]

{Lattice QCD \cite{bec}} &\hspace{0.6in}{$-$} &\hspace{0.5in}{$-$} &\hspace{0.5in}{$1.10\pm0.02$} &\hspace{0.5in}{$1.11\pm0.03$} \\[1 ex]

{BS \cite{wang2}} &\hspace{0.6in}{$-$} &\hspace{0.5in}{$-$} &\hspace{0.5in}{$1.01$} &\hspace{0.5in}{$-$} \\[1 ex]

{BS \cite{cvetic, wang3}} &\hspace{0.6in}{$-$} &\hspace{0.5in}{$-$} &\hspace{0.5in}{$1.08\pm0.01$} &\hspace{0.5in}{$1.10\pm0.06$} \\[1 ex]

{RQM \cite{hwang2}} &\hspace{0.6in}{$1.21\pm0.02$} &\hspace{0.5in}{$1.17\pm0.02$} &\hspace{0.5in}{$1.14\pm0.01$} &\hspace{0.5in}{$-$} \\[1 ex]

{RQM \cite{capstick}} &\hspace{0.6in}{$-$} &\hspace{0.5in}{$-$} &\hspace{0.5in}{$1.21\pm0.13$} &\hspace{0.5in}{$-$} \\[1 ex]

{RQM \cite{ebert}} &\hspace{0.6in}{$1.32$} &\hspace{0.5in}{$1.18$} &\hspace{0.5in}{$1.15$} &\hspace{0.5in}{$1.02$} \\[1 ex]

{NRQM \cite{yazar2}} &\hspace{0.6in}{$0.97$} &\hspace{0.5in}{$0.97$} &\hspace{0.5in}{$1.11$} &\hspace{0.5in}{$1.12$} \\[1 ex]

{NRQM \cite{yazar3}} &\hspace{0.6in}{$0.96$} &\hspace{0.5in}{$0.97$} &\hspace{0.5in}{$1.07$} &\hspace{0.5in}{$1.08$} \\[1 ex]

{NRQM \cite{shady}} &\hspace{0.6in}{$1.32$} &\hspace{0.5in}{$1.24$} &\hspace{0.5in}{$1.14$} &\hspace{0.5in}{$1.07$} \\[1 ex]

{NRQM \cite{hass}} &\hspace{0.6in}{$-$} &\hspace{0.5in}{$-$} &\hspace{0.5in}{$1.20$} &\hspace{0.5in}{$-$} \\[1 ex]

\hline\hline
\end{tabular}
\end{table}

\begin{table}[h]
\centering
\caption{ Pseudoscalar, longitudinally and transversely polarized vector $B$ meson decay constants (in units of MeV) in the present work and their comparison with the available experimental data and other theoretical model predictions.}
\vspace{0.2 cm}
\label{tab8}
\begin{tabular}{p{1.35 in}m{0.85 in}p{0.75 in}p{0.75 in}p{0.85 in}p{0.75 in}p{0.75 in}} 
\hline\hline\\[-1.5 ex]
\multicolumn{1}{c}{} &\multicolumn{1}{c}{$f_{B}$} & \multicolumn{1}{c}{$f_{B^*}$} &\multicolumn{1}{c}{$f_{B^*}^{\perp}$} &\multicolumn{1}{c}{$f_{B_s}$} &\multicolumn{1}{c}{$f_{B_s^*}$} &\multicolumn{1}{c}{$f_{B_s^*}^{\perp}$} \\[1 ex]
\hline\\[-1.5 ex]

{Present work} &\multicolumn{1}{c}{$163^{+21-4}_{-20+4}$} &\multicolumn{1}{c}{$172^{+24-6}_{-23+6}$} &\multicolumn{1}{c}{$165^{+22-5}_{-21+5}$} 
&\multicolumn{1}{c}{$184^{+23- 4}_{-23+4}$} &\multicolumn{1}{c}{$194^{+26-6}_{-25+7}$} &\multicolumn{1}{c}{$187^{+24-5}_{-24+6}$}\\[1 ex]

{Exp. \cite{tana}} &\multicolumn{1}{c}{$188\pm 25$} &\quad \quad{$-$} &\quad \quad{$-$} &\quad \quad{$-$} &\quad \quad{$-$} &\quad \quad{$-$}\\[1 ex]

{LFQM, Lin \cite{choi1}} &\multicolumn{1}{c}{171} &\multicolumn{1}{c}{186} &\multicolumn{1}{c}{$-$} &\multicolumn{1}{c}{205} &\multicolumn{1}{c}{220} &\multicolumn{1}{c}{$-$}\\[1 ex]

{LFQM, HO \cite{choi1}} &\multicolumn{1}{c}{161} &\multicolumn{1}{c}{173} &\multicolumn{1}{c}{$-$} &\multicolumn{1}{c}{208} &\multicolumn{1}{c}{223} &\multicolumn{1}{c}{$-$}\\[1 ex]

{LFQM, Lin \cite{choi2}} &\multicolumn{1}{c}{181} &\multicolumn{1}{c}{188} &\multicolumn{1}{c}{$-$} &\multicolumn{1}{c}{205} &\multicolumn{1}{c}{216} &\multicolumn{1}{c}{$-$}\\[1 ex]

\multirow{2}{5em}{LFQM \cite{hwang1}} &\multicolumn{1}{c}{$-$} &\multicolumn{1}{c}{$225\pm38$} &\multicolumn{1}{c}{$214\pm34$} &\multicolumn{1}{c}{$281\pm54$} &\multicolumn{1}{c}{$313\pm67$} &\multicolumn{1}{c}{$297\pm61$} \\
&\multicolumn{1}{c}{} &\multicolumn{1}{c}{$249^{+44}_{-42}$} &\multicolumn{1}{c}{$226\pm37$} &\multicolumn{1}{c}{$270\pm47$} &\multicolumn{1}{c}{$335\pm68$} &\multicolumn{1}{c}{$302\pm58$} \\[1 ex]

{LFQM \cite{chao1}} &\multicolumn{1}{c}{$-$} &\multicolumn{1}{c}{$201.9^{+43.2}_{-41.4}$} &\multicolumn{1}{c}{$-$} &\multicolumn{1}{c}{$-$} &\multicolumn{1}{c}{$244.2\pm7.0$} &\multicolumn{1}{c}{$-$}\\[1 ex]

{LFHQCD \cite{dosch}} &\multicolumn{1}{c}{$194$} &\multicolumn{1}{c}{$-$} &\multicolumn{1}{c}{$-$} &\multicolumn{1}{c}{$229$} &\multicolumn{1}{c}{$-$} &\multicolumn{1}{c}{$-$}\\[1 ex]

{LFHQCD \cite{chang1}} &\multicolumn{1}{c}{$191.7^{+7.9}_{-6.5}$} &\multicolumn{1}{c}{$-$} &\multicolumn{1}{c}{$-$} &\multicolumn{1}{c}{$225.4^{+7.9}_{-5.3}$} &\multicolumn{1}{c}{$-$} &\multicolumn{1}{c}{$-$}\\[1 ex]

{SR \cite{narison}} &\multicolumn{1}{c}{$204\pm5.1$} &\multicolumn{1}{c}{$210\pm6$} &\multicolumn{1}{c}{$-$} &\multicolumn{1}{c}{$234.5\pm4.4$} &\multicolumn{1}{c}{$221\pm7$} &\multicolumn{1}{c}{$-$}\\[1 ex]

{SR \cite{wang1}} &\multicolumn{1}{c}{$194\pm15$} &\multicolumn{1}{c}{$213\pm18$} &\multicolumn{1}{c}{$-$} &\multicolumn{1}{c}{$231\pm16$} &\multicolumn{1}{c}{$255\pm19$} &\multicolumn{1}{c}{$-$}\\[1 ex]

{SR \cite{gel}} &\multicolumn{1}{c}{$207^{+17}_{-9}$} &\multicolumn{1}{c}{$210^{+10}_{-12}$} &\multicolumn{1}{c}{$-$} &\multicolumn{1}{c}{$242^{+17}_{-12}$} &\multicolumn{1}{c}{$251^{+14}_{-16}$} &\multicolumn{1}{c}{$-$}\\[1 ex]

{Lattice QCD \cite{aoki}} &\multicolumn{1}{c}{$187.1\pm4.2$} &\multicolumn{1}{c}{$-$} &\multicolumn{1}{c}{$-$} &\multicolumn{1}{c}{$227.2\pm3.4$} &\multicolumn{1}{c}{$-$} &\multicolumn{1}{c}{$-$}\\[1 ex]

{Lattice QCD \cite{bec}} &\multicolumn{1}{c}{$179\pm18^{+34}_{-9}$} &\multicolumn{1}{c}{$196\pm24^{+39}_{-2}$} &\multicolumn{1}{c}{$-$} &\multicolumn{1}{c}{$204\pm16^{+36}_{-0}$} &\multicolumn{1}{c}{$229\pm20^{+41}_{-16}$} &\multicolumn{1}{c}{$-$}\\[1 ex]

{Lattice QCD \cite{na1}} &\multicolumn{1}{c}{$191\pm9$} &\multicolumn{1}{c}{$-$} &\multicolumn{1}{c}{$-$} &\multicolumn{1}{c}{$228\pm10$} &\multicolumn{1}{c}{$-$} &\multicolumn{1}{c}{$-$}\\[1 ex]

{BS \cite{wang2}} &\multicolumn{1}{c}{$193$} &\multicolumn{1}{c}{$-$} &\multicolumn{1}{c}{$-$} &\multicolumn{1}{c}{$195$} &\multicolumn{1}{c}{$-$} &\multicolumn{1}{c}{$-$}\\[1 ex]

{BS \cite{cvetic, wang3}} &\multicolumn{1}{c}{$196\pm29$} &\multicolumn{1}{c}{$238\pm18$} &\multicolumn{1}{c}{$-$} &\multicolumn{1}{c}{$216\pm32$} &\multicolumn{1}{c}{$272\pm20$} &\multicolumn{1}{c}{$-$}\\[1 ex]

{RQM \cite{hwang2}} &\multicolumn{1}{c}{$231\pm9$} &\multicolumn{1}{c}{$252\pm10$} &\multicolumn{1}{c}{$-$} &\multicolumn{1}{c}{$266\pm10$} &\multicolumn{1}{c}{$289\pm11$} &\multicolumn{1}{c}{$-$}\\[1 ex]

{RQM \cite{capstick}} &\multicolumn{1}{c}{$155\pm15$} &\multicolumn{1}{c}{$-$} &\multicolumn{1}{c}{$-$} &\multicolumn{1}{c}{$210\pm20$} &\multicolumn{1}{c}{$-$} &\multicolumn{1}{c}{$-$}\\[1 ex]

{RQM \cite{ebert}} &\multicolumn{1}{c}{$189$} &\multicolumn{1}{c}{$219$} &\multicolumn{1}{c}{$-$} &\multicolumn{1}{c}{$218$} &\multicolumn{1}{c}{$251$} &\multicolumn{1}{c}{$-$}\\[1 ex]

{NRQM \cite{yazar1}} &\multicolumn{1}{c}{$243.64$} &\multicolumn{1}{c}{$242.37$} &\multicolumn{1}{c}{$-$} &\multicolumn{1}{c}{$179.21$} &\multicolumn{1}{c}{$178.82$} &\multicolumn{1}{c}{$-$}\\[1 ex]

{NRQM \cite{yazar3}} &\multicolumn{1}{c}{$235.9$} &\multicolumn{1}{c}{$234.7$} &\multicolumn{1}{c}{$-$} &\multicolumn{1}{c}{$245.1$} &\multicolumn{1}{c}{$244.2$} &\multicolumn{1}{c}{$-$}\\[1 ex]

{NRQM \cite{shady}} &\multicolumn{1}{c}{$147$} &\multicolumn{1}{c}{$196$} &\multicolumn{1}{c}{$-$} &\multicolumn{1}{c}{$174$} &\multicolumn{1}{c}{$216$} &\multicolumn{1}{c}{$-$}\\[1 ex]

{NRQM \cite{hass}} &\multicolumn{1}{c}{$149$} &\multicolumn{1}{c}{$-$} &\multicolumn{1}{c}{$-$} &\multicolumn{1}{c}{$187$} &\multicolumn{1}{c}{$-$} &\multicolumn{1}{c}{$-$}\\[1 ex]

\hline\hline
\end{tabular}
\end{table}

\begin{table}[h]
\centering
\caption{ Ratio of the decay constants for $(B, B_s, B^*, B^*_s)$ mesons compared with the available experimental data and other theoretical model calculations.}
\vspace{0.2 cm}
\label{tab9}
\begin{tabular}{lcccc}
\hline\hline\\[-1.5 ex]
\multicolumn{1}{c}{} &\hspace{0.6in}{$f_{B^*}/f_{B}$} & \hspace{0.5in}{$f_{B_s^*}/f_{B_s}$} &\hspace{0.5in}{$f_{B_s}/f_{B}$} &\hspace{0.5in}{$f_{B_s^*}/f_{B^*}$} \\[1 ex]
\hline\\[-1.5 ex]

{Present work} &\hspace{0.2in}{$1.06^{+0.010-0.011}_{-0.013+0.011}$} 
&\hspace{0.2in}{$1.05^{+0.008-0.010}_{-0.005+0.015}$} 
&\hspace{0.2in}{$1.13^{-0.004+0.003}_{-0.003-0.003}$} 
&\hspace{0.2in}{$1.13^{-0.005+0.005}_{+0.006+0.001}$} \\[1 ex]

{LFQM, Lin \cite{choi1}} &\hspace{0.6in}{1.09} &\hspace{0.5in}{1.07} &\hspace{0.5in}{1.20} &\hspace{0.5in}{1.18} \\[1 ex]

{LFQM, HO \cite{choi1}} &\hspace{0.6in}{1.07} &\hspace{0.5in}{1.07} &\hspace{0.5in}{1.29} &\hspace{0.5in}{1.29} \\[1 ex]

{LFQM, Lin \cite{choi2}} &\hspace{0.6in}{1.04} &\hspace{0.5in}{1.05} &\hspace{0.5in}{1.13} &\hspace{0.5in}{1.15} \\[1 ex]

\multirow{2}{5em}{LFQM \cite{hwang1}} &\hspace{0.6in}{$1.10\pm0.02$} &\hspace{0.5in}{$1.11\pm0.03$} &\hspace{0.5in}{$1.38\pm0.07$} &\hspace{0.5in}{$1.39\pm0.08$} \\
&\hspace{0.6in}{$1.22\pm0.03$} &\hspace{0.5in}{$1.24\pm0.05$} &\hspace{0.5in}{$1.32\pm0.08$} &\hspace{0.5in}{$1.35\pm0.08$} \\[1 ex]

{LFQM \cite{chao1}} &\hspace{0.6in}{$1.09^{+0.31}_{-0.30}$} &\hspace{0.5in}{$1.09\pm0.04$} &\hspace{0.5in}{$-$} &\hspace{0.5in}{$-$} \\[1 ex]

{LFHQCD \cite{dosch}} &\hspace{0.6in}{$-$} &\hspace{0.5in}{$-$} &\hspace{0.5in}{1.18} &\hspace{0.5in}{$-$} \\[1 ex]

{LFHQCD \cite{chang1}} &\hspace{0.6in}{$-$} &\hspace{0.5in}{$-$} &\hspace{0.5in}{$1.176^{+0.056}_{-0.053}$} &\hspace{0.5in}{$-$} \\[1 ex]

{SR \cite{narison}} &\hspace{0.6in}{$1.020\pm11$} &\hspace{0.5in}{$-$} &\hspace{0.5in}{$1.154\pm21$} &\hspace{0.5in}{$1.064\pm10$} \\[1 ex]

{SR \cite{wang1}} &\hspace{0.6in}{$-$} &\hspace{0.5in}{$-$} &\hspace{0.5in}{$1.19\pm0.10$} &\hspace{0.5in}{$-$} \\[1 ex]

{SR \cite{gel}} &\hspace{0.6in}{$1.02^{+0.02}_{-0.09}$} &\hspace{0.5in}{$1.04^{+0.01}_{-0.08}$} &\hspace{0.5in}{$1.17^{+0.03}_{-0.04}$} &\hspace{0.5in}{$1.20\pm0.04$} \\[1 ex]

{Lattice QCD \cite{aoki}} &\hspace{0.6in}{$-$} &\hspace{0.5in}{$-$} &\hspace{0.5in}{$1.215\pm0.007$} &\hspace{0.5in}{$-$} \\[1 ex]

{Lattice QCD \cite{bec}} &\hspace{0.6in}{} &\hspace{0.5in}{} &\hspace{0.5in}{$1.14\pm0.03^{+1}_{-1}$} &\hspace{0.5in}{$1.17\pm0.04^{+1}_{-3}$} \\[1 ex]

{Lattice QCD \cite{na1}} &\hspace{0.6in}{$-$} &\hspace{0.5in}{$-$} &\hspace{0.5in}{$1.188\pm18$} &\hspace{0.5in}{$-$} \\[1 ex]

{BS \cite{wang2}} &\hspace{0.6in}{$-$} &\hspace{0.5in}{$-$} &\hspace{0.5in}{$1.01$} &\hspace{0.5in}{$-$} \\[1 ex]

{BS \cite{cvetic, wang3}} &\hspace{0.6in}{$-$} &\hspace{0.5in}{$-$} &\hspace{0.5in}{$1.10\pm0.01$} &\hspace{0.5in}{$1.14\pm0.08$} \\[1 ex]

{RQM \cite{hwang2}} &\hspace{0.6in}{$1.09\pm0.01$} &\hspace{0.5in}{$1.09\pm0.01$} &\hspace{0.5in}{$1.15\pm0.01$} &\hspace{0.5in}{$-$} \\[1 ex]

{RQM \cite{capstick}} &\hspace{0.6in}{$-$} &\hspace{0.5in}{$-$} &\hspace{0.5in}{$1.35\pm0.18$} &\hspace{0.5in}{$-$} \\[1 ex]

{RQM \cite{ebert}} &\hspace{0.6in}{$1.16$} &\hspace{0.5in}{$1.15$} &\hspace{0.5in}{$1.15$} &\hspace{0.5in}{$1.15$} \\[1 ex]

{NRQM \cite{yazar1}} &\hspace{0.6in}{$0.99$} &\hspace{0.5in}{$1.00$} &\hspace{0.5in}{$0.74$} &\hspace{0.5in}{$0.74$} \\[1 ex]

{NRQM \cite{yazar3}} &\hspace{0.6in}{$0.99$} &\hspace{0.5in}{$1.00$} &\hspace{0.5in}{$1.04$} &\hspace{0.5in}{$1.04$} \\[1 ex]

{NRQM \cite{shady}} &\hspace{0.6in}{$1.33$} &\hspace{0.5in}{$1.24$} &\hspace{0.5in}{$1.18$} &\hspace{0.5in}{$1.10$} \\[1 ex]

{NRQM \cite{hass}} &\hspace{0.6in}{$-$} &\hspace{0.5in}{$-$} &\hspace{0.5in}{$1.26$} &\hspace{0.5in}{$-$} \\[1 ex]

\hline\hline
\end{tabular}
\end{table}

\begin{table}[h]
\centering
\caption{First six $\xi$-moments of ($D_{(s)}, D^*_{(s)}$) mesons.}
\vspace{0.2 cm}
\label{tab10}
\begin{tabular}{lcccccc}
\hline\hline\\[-1.5 ex]
\multicolumn{1}{c}{} &\hspace{0.6in}{$\langle \xi^{1} \rangle$} &\hspace{0.5in}{$\langle \xi^{2} \rangle$} &\hspace{0.5in}{$\langle \xi^{3} \rangle$} &\hspace{0.5in}{$\langle \xi^{4} \rangle$} &\hspace{0.5in}{$\langle \xi^{5} \rangle$} &\hspace{0.5in}{$\langle \xi^{6} \rangle$} \\[1 ex]
\hline\\[-1.5 ex]

{$D$} &\hspace{0.6in}{0.325} &\hspace{0.5in}{0.218} &\hspace{0.5in}{0.142} &\hspace{0.5in}{0.106} &\hspace{0.5in}{0.081} &\hspace{0.5in}{0.065} \\[1 ex]

{$D^*$} &\hspace{0.6in}{0.356} &\hspace{0.5in}{0.227} &\hspace{0.5in}{0.149} &\hspace{0.5in}{0.110} &\hspace{0.5in}{0.083} &\hspace{0.5in}{0.066} \\[1 ex]

{$D^*_\perp$} &\hspace{0.6in}{0.351} &\hspace{0.5in}{0.226} &\hspace{0.5in}{0.149} &\hspace{0.5in}{0.110} &\hspace{0.5in}{0.084} &\hspace{0.5in}{0.066} \\[1 ex]

{$D_s$} &\hspace{0.6in}{0.311} &\hspace{0.5in}{0.202} &\hspace{0.5in}{0.125} &\hspace{0.5in}{0.089} &\hspace{0.5in}{0.065} &\hspace{0.5in}{0.049} \\[1 ex]

{$D_s^*$} &\hspace{0.6in}{0.323} &\hspace{0.5in}{0.202} &\hspace{0.5in}{0.125} &\hspace{0.5in}{0.088} &\hspace{0.5in}{0.063} &\hspace{0.5in}{0.048} \\[1 ex]

{$D^*_{s_\perp}$} &\hspace{0.6in}{0.321} &\hspace{0.5in}{0.203} &\hspace{0.5in}{0.126} &\hspace{0.5in}{0.088} &\hspace{0.5in}{0.064} &\hspace{0.5in}{0.049} \\[1 ex]

\hline\hline
\end{tabular}
\end{table}

\begin{table}[h]
\centering
\caption{First six $\xi$-moments of ($B_{(s)}, B^*_{(s)}$) mesons.}
\vspace{0.2 cm}
\label{tab11}
\begin{tabular}{lcccccc}
\hline\hline\\[-1.5 ex]
\multicolumn{1}{c}{} &\hspace{0.6in}{$\langle \xi^{1} \rangle$} &\hspace{0.5in}{$\langle \xi^{2} \rangle$} &\hspace{0.5in}{$\langle \xi^{3} \rangle$} &\hspace{0.5in}{$\langle \xi^{4} \rangle$} &\hspace{0.5in}{$\langle \xi^{5} \rangle$} &\hspace{0.5in}{$\langle \xi^{6} \rangle$} \\[1 ex]
\hline\\[-1.5 ex]

{$B$} &\hspace{0.6in}{0.665} &\hspace{0.5in}{0.471} &\hspace{0.5in}{0.351} &\hspace{0.5in}{0.273} &\hspace{0.5in}{0.219} &\hspace{0.5in}{0.180} \\[1 ex]

{$B^*$} &\hspace{0.6in}{0.672} &\hspace{0.5in}{0.480} &\hspace{0.5in}{0.360} &\hspace{0.5in}{0.280} &\hspace{0.5in}{0.225} &\hspace{0.5in}{0.185} \\[1 ex]

{$B^*_\perp$} &\hspace{0.6in}{0.672} &\hspace{0.5in}{0.480} &\hspace{0.5in}{0.359} &\hspace{0.5in}{0.280} &\hspace{0.5in}{0.224} &\hspace{0.5in}{0.185} \\[1 ex]

{$B_s$} &\hspace{0.6in}{0.651} &\hspace{0.5in}{0.452} &\hspace{0.5in}{0.331} &\hspace{0.5in}{0.253} &\hspace{0.5in}{0.199} &\hspace{0.5in}{0.161} \\[1 ex]

{$B_s^*$} &\hspace{0.6in}{0.652} &\hspace{0.5in}{0.455} &\hspace{0.5in}{0.334} &\hspace{0.5in}{0.254} &\hspace{0.5in}{0.200} &\hspace{0.5in}{0.161} \\[1 ex]

{$B^*_{s_\perp}$} &\hspace{0.6in}{0.653} &\hspace{0.5in}{0.456} &\hspace{0.5in}{0.334} &\hspace{0.5in}{0.255} &\hspace{0.5in}{0.201} &\hspace{0.5in}{0.162} \\[1 ex]

\hline\hline
\end{tabular}
\end{table}

\newpage
\appendix*
\section{ANALYTIC FORMULA OF THE MASS EIGENBALUES OF THE GROUND STATE PSEUDOSCALAR AND VECTOR MESONS OBTAINED USING EXPONENTIAL-TYPE POTENTIAL}
\label{app1}
\vspace{-4.5cm}
\begin{eqnarray}
\label{eqn:32}
M_{q \bar{q}} &=& a + \frac{b}{3840 \sqrt{\pi} \beta^{10}} \bigg\{80 \bigg(24 c_1^2 \Big(2 \alpha \beta + e^{\frac{\alpha^2}{4 \beta^2}} \sqrt{\pi} \big(\alpha^2 + 2 \beta^2 \big) \Big) \beta^4 - 4 \sqrt{6} c_1 c_2 \alpha \Big(2 \beta \big(\alpha^2 + 4 \beta^2 \big) \nonumber \\
&& + e^{\frac{\alpha^2}{4 \beta^2}} \sqrt{\pi} \alpha  \big(\alpha^2 + 6 \beta^2 \big)\Big) \beta^2 + c_2^2 \Big(2 \alpha \beta \big(\alpha^4 + 16 \alpha^2 \beta^2 + 48 \beta^4 \big) + e^{\frac{\alpha^2}{4 \beta^2}} \sqrt{\pi} \big(\alpha^6 + 18 \alpha^4 \beta^2 \nonumber \\
&& + 72 \alpha^2 \beta^4 + 48 \beta^6 \big) \Big) \bigg) \beta^4 - 8 \sqrt{5} c_3 \alpha \bigg(2 \sqrt{6} c_1 \Big(2 \beta \big(-\alpha^4 - 8 \alpha^2 \beta^2 + 8 \beta^4 \big) - e^{\frac{\alpha^2}{4 \beta^2}} \sqrt{\pi} \alpha^3 \nonumber \\
&& \big(\alpha^2 + 10 \beta^2 \big)\Big) \beta^2 + c_2 \Big(2 \beta \big(\alpha^6 + 28 \alpha^4 \beta^2 + 192 \alpha^2 \beta^4 + 240 \beta^6 \big) + e^{\frac{\alpha^2}{4 \beta^2}} \sqrt{\pi} \alpha  \big(\alpha^6 + 30 \alpha^4 \beta^2 \nonumber \\
&& + 240 \alpha^2 \beta^4 + 480 \beta^6 \big)\Big)\bigg) \beta^2 + c_3^2 \bigg(2 \alpha \beta \Big(\alpha^8 + 48 \alpha^6 \beta^2 + 712 \alpha^4 \beta^4 + 3680 \alpha^2 \beta^6 + 5280 \beta^8 \Big)\nonumber \\
&& + e^{\frac{\alpha^2}{4 \beta^2}} \sqrt{\pi} \Big(\alpha^{10} + 50 \alpha^8 \beta^2 + 800 \alpha^6 \beta^4 + 4800 \alpha^4 \beta^6 + 9600 \alpha^2 \beta^8 + 3840 \beta^{10}\Big)\bigg)\bigg\} \nonumber \\
&& + \frac{b e^{\frac{\alpha^2}{4 \beta^2}}}{3840 \beta^{10}} \bigg\{80 \Big(24 c_1^2 \big(\alpha^2 + 2 \beta^2 \big) \beta^4 - 4 \sqrt{6} c_1 c_2 \alpha^2 \big(\alpha^2 + 6 \beta^2 \big) \beta^2 + c_2^2 \big(\alpha^6 + 18 \beta^2 \alpha^4 \nonumber \\
&& + 72 \beta^4 \alpha^2 + 48 \beta^6 \big)\Big) \beta^4 - 8 \sqrt{5} c_3 \alpha^2 \Big(c_2 \big(\alpha^6 + 30 \alpha^4 \beta^2 + 240 \alpha^2 \beta^4 + 480 \beta^6 \big) - 2 \sqrt{6} c_1 \alpha^2 \beta^2 \nonumber \\
&& \big(\alpha^2 + 10 \beta^2 \big)\Big) \beta^2 + c_3^2 \Big(\alpha^{10}  + 50 \alpha^8 \beta^2 + 800 \alpha^6 \beta^4 + 4800 \alpha^4 \beta^6 + 9600 \alpha^2 \beta^8 + 3840 \beta^{10} \Big)\bigg\} \nonumber \\
&& {\rm erf}\bigg(\frac{\alpha}{2 \beta} \bigg) + \frac{1}{120 \sqrt{\pi} \beta^5} \sum_{i = q, \bar{q}} \Bigg\{15 \sqrt{\pi} \Bigg(63 c_3^2 U\bigg(-\frac{1}{2}, -5, \frac{m_i^2}{\beta^2}\bigg) + 4 \bigg(7 c_3 \Big(\sqrt{5} c_2 - 5 c_3 \Big) \nonumber \\
&& U\bigg(-\frac{1}{2}, -4, \frac{m_i^2}{\beta^2}\bigg) + \Big(-6 c_2^2 + 9 \sqrt{5} c_2 c_3 - 15 c_3^2 + 2 \sqrt{6} c_1 \big(c_2 - \sqrt{5} c_3 \big)\Big) \nonumber \\
&& U\bigg(-\frac{1}{2}, -2, \frac{m_i^2}{\beta^2}\bigg)\bigg)\Bigg) \beta^6 - 4 m_i^4 \big(m_i^2 - 3 \beta^2 \big) e^{\frac{m_i^2}{2 \beta^2}} \Big(10 c_2^2 - 26 \sqrt{5} c_2 c_3 + c_3 \big(2 \sqrt{30} c_1 \nonumber \\
&& + 65 c_3 \big)\Big) K_2\bigg(\frac{m_i^2}{2 \beta^2}\bigg) + m_i^2 e^{\frac{m_i^2}{2 \beta^2}} \bigg(120 c_1^2 \beta^4 + 20 c_2^2 \big(2 m_i^4 + 9 \beta^4 \big) + 4 \sqrt{30} c_1 c_3 \big(2 m_i^4 \nonumber \\
&& + 15 \beta^4 \big) + 5 c_3^2 \big(52 m_i^4 + 45 \beta^4 \big) - 4 c_2 \Big(30 \sqrt{6} c_1 \beta^4 + \sqrt{5} c_3 \big(26 m_i^4 + 45 \beta^4 \big)\Big)\bigg) K_1\bigg(\frac{m_i^2}{2 \beta^2}\bigg)\bigg\} \nonumber \\
&& -\frac{\big(120 c_1^2 + 100 c_2^2 + 89 c_3^2 + 40 \sqrt{6} c_1 c_2 + 44 \sqrt{5} c_2 c_3 + 12 \sqrt{30} c_1 c_3 \big) \kappa \beta}{45 \sqrt{\pi}} \nonumber \\
&& + \frac{4 \kappa \beta^3 \langle\bf{{S}_{q}}\cdot\bf{{S}_{\bar{q}}}\rangle}{9 {m_q} {m_{\bar{q}}} \sqrt{\pi } \left(\beta^2 + \sigma^2\right)^{11/2}}
\bigg\{8 c_1^2(\beta^2 + \sigma^2)^4 + 4 c_2^2(2 \beta^4 + 3 \sigma^4)(\beta^2 + \sigma^2)^2 + 4 \sqrt{6} c_1 \sigma^2 \nonumber \\
&& \big(\sqrt{5} c_3 \sigma^2 + 2 c_2 (\beta^2 + \sigma^2)\big) (\beta^2 + \sigma^2)^2 + 4 \sqrt{5} c_2 c_3 \sigma^2 (4 \beta^4 + 3 \sigma^4) (\beta^2 + \sigma^2) \nonumber \\
&& + c_3^2 (8 \beta ^8 + 40 \sigma^4 \beta^4 + 15 \sigma^8)\bigg\} \sigma^3,
\end{eqnarray}
where $K_{1 (2)}$ represents the modified Bessel function of the second kind and $U (a, b, z )$ represents the Tricomi's (confluent hypergeometric) function.

\end{document}